\documentclass[conference,compsoc]{IEEEtran}
\usepackage{adjustbox}
\usepackage{amssymb}
\usepackage{booktabs}
\usepackage{cite}
\usepackage{color}
\usepackage{enumitem}
\usepackage{float}
\usepackage[T1]{fontenc}
\usepackage{graphicx}
\usepackage[utf8]{inputenc}
\usepackage{multicol}
\usepackage{multirow}
\usepackage{subfig}
\usepackage{pifont}
\usepackage[normalem]{ulem}
\usepackage{tabularx}
\usepackage{threeparttable}
\usepackage{stfloats}
\usepackage{comment}
\usepackage{hyperref}

\newcommand{\cmark}{\ding{51}}%
\newcommand{\xmark}{\ding{55}}%

\usepackage[camera]{dtrt}
\usepackage[capitalize]{cleveref}

\definecolor{camel}{rgb}{0.76, 0.6, 0.42}

\usepackage{array}
\newcolumntype{H}{>{\setbox0=\hbox\bgroup}c<{\egroup}@{}}

\newcommand{\myparagraphit}[1]{{\smallskip\textit{#1}}}
\newcommand{\MEVAP}{MEV auction platform\xspace}
\newcommand{\MEVAPs}{MEV auction platforms\xspace}
\usepackage{xspace}
\newcommand{\numberofprojects}{{30}\xspace}

\begin{document}

\title{SoK: MEV Countermeasures: Theory and Practice}

\author{\IEEEauthorblockN{Sen Yang\IEEEauthorrefmark{1}, Fan Zhang\IEEEauthorrefmark{1}, Ken Huang\IEEEauthorrefmark{2}, Xi Chen\IEEEauthorrefmark{3}, Youwei Yang\IEEEauthorrefmark{4}, and Feng Zhu\IEEEauthorrefmark{5}}
\IEEEauthorblockA{\IEEEauthorrefmark{1}Department of Computer Science, Yale University}
\IEEEauthorblockA{\IEEEauthorrefmark{2}DistributedApps}
\IEEEauthorblockA{\IEEEauthorrefmark{3}Stern School of Business, New York University}
\IEEEauthorblockA{\IEEEauthorrefmark{4}Bit Mining Limited}
\IEEEauthorblockA{\IEEEauthorrefmark{5}Harvard Business School}}

\date{\today}
\maketitle              %
\thispagestyle{plain}
\begin{abstract}

Blockchains offer strong security guarantees, but they cannot protect the ordering of transactions.
Powerful players, such as miners, sequencers, and sophisticated bots, can reap significant profits by selectively including, excluding, or re-ordering user transactions.
Such profits are called Miner/Maximal Extractable Value or MEV.
MEV bears profound implications for blockchain security and decentralization.
While numerous countermeasures have been proposed, there is no agreement on the best solution. 
Moreover, solutions developed in academic literature differ quite drastically from what is widely adopted by practitioners.
For these reasons, this paper systematizes the knowledge of the theory and practice of MEV countermeasures.
The contribution is twofold.
First, we present a comprehensive taxonomy of \numberofprojects proposed MEV countermeasures, covering four different technical directions.
Secondly, we empirically studied the most popular MEV-auction-based solution with rich blockchain and mempool data. We also present the Mempool Guru system, a public service system that collects, persists, and analyzes the Ethereum mempool data for research.
In addition to gaining insights into \MEVAPs' real-world operations, our study shed light on the prevalent censorship by \MEVAPs as a result of the recent OFAC sanction, and its implication on blockchain properties.
\end{abstract}

\section{Introduction}
\label{sec:intro}

Smart contracts are autonomous programs running on top of a blockchain. They achieve strong security properties (integrity, transparency, censorship-resistance, etc) that centralized systems cannot offer and have cultivated a trillion-dollar ecosystem spanning finance products (e.g., stablecoins, lending), markets and exchanges (e.g., automated market makers), digital assets (e.g., NFTs) and more.

However, blockchains cannot protect the {\em ordering of transactions}~\cite{daian2020flash}.
The ability to manipulate transaction ordering is immense power.
For instance, if Alice can execute her trades before Bob's {\em ex-post} (i.e., after having observed Bob's trades), she can frontrun~\cite{markham1988front} Bob and reap a profit. 
Daian et al.~\cite{daian2020flash} coined the term Miner/Maximal Extractable Value (MEV) to denote the profits that powerful players (with miners being the most powerful) can gain (or {\em extract}) from selectively including, excluding, or re-ordering user transactions.

MEV bears profound implications for the security and decentralization of blockchain systems. Some MEV is at the expense of regular users (e.g., sandwich attacks), which should be prevented.
Some MEV is benign (e.g., arbitrage profits provide incentives for price discovery and price synchronization cross exchanges), but uncoordinated extraction of it can cause network congestion and high transaction fees~\cite{daian2020flash}. Moreover, when MEV dominates block rewards (which they already do quite often in Ethereum), blockchain consensus could be destabilized~\cite{carlsten2016instability}. Last but not least, MEV can lead to centralization because finding sophisticated MEV opportunities require significant resources (money and talent) only big players can afford~\cite{pbs-vitalik-post}.

Academia and the industry have attempted countermeasures from several different directions.
At the highest level of abstraction, two schools of ideas were explored: 1) to facilitate MEV extraction so that the process is efficient, decentralized and transparent~\cite{flashbots,mevboost,pbs}, and 2) to stop MEV extraction by ordering transactions in a way that renders order manipulation infeasible (e.g., \cite{kelkar2020order,kelkar2021themis,cachin2022quick}).
As readers can notice, the two directions aim to lead to different and even possibly incompatible futures. There is currently no consensus on the best countermeasures. 
Moreover, while the second approach is better explored by academics, it is the first approach that is widely adopted in practice thus far.

\parhead{Our contribution.} This paper systematizes the knowledge of MEV countermeasures, covering both academically proposed solutions and the popular choice of  practitioners. Thus the contribution is twofold. First, we present a comprehensive taxonomy of recently proposed MEV countermeasures, covering four different technical directions. Second, we empirically study the most popular solutions in practice, with rich blockchain and mempool data.

\parhead{A taxonomy of MEV countermeasures.}
We selected \numberofprojects recent projects and papers that proposed representative MEV countermeasures, from the following four categories.

\myparagraphit{I: \MEVAPs.} This class of solutions builds the facility to make MEV extraction efficient, decentralized, and transparent. We call entities who actively extract MEV as {\em MEV searchers}.  Most \MEVAPs guarantee the privacy and atomicity of searchers' transactions. The former protects MEV searchers from other searchers. The latter is important for MEV extraction that involves executing multiple transactions in a particular order. Atomicity ensures that either all of the transactions are executed in the desired order or none of them is, but never partially. Two guarantees together drastically reduce the risk of MEV extraction.

\myparagraphit{II: Time-based order fairness.} Fundamentally, state machine replication protocols (which blockchains realize) can be extended to enforce an extended validity property that the ordering of transactions must satisfy. Different notions of {\em order fairness} are defined. Kelkar et al.~\cite{kelkar2020order,kelkar2021themis} proposed receive-order fairness which requires if a majority of nodes receive $T_1$ before $T_2$, then the final blockchain should respect that order. Order linearizability~\cite{zhang2020byzantine}, $\kappa$-differential order-fairness~\cite{cachin2022quick}, timed-relative fairness~\cite{kursawe2020wendy} are related notions. Enforcing order fairness on user transactions can prevent MEV caused by {\em ex-post} order manipulation. 

\myparagraphit{III: Content-agnostic ordering.} A particular (weaker) fairness property that received a lot of attention is content-agnostic ordering, which means the order of transactions is determined independently of transaction content.
The high-level idea is to have users first commit to transactions, and reveal them after the ordering has been determined.
Content-agnostic ordering is weaker than receive-order fairness because it permits metadata leakage (e.g., many schemes leak the sender so that a fee can be charged to prevent DoS attacks) and content-agnostic frontrunning (e.g., when the attacker just wants to place her transaction before others). While weaker, content-agnostic ordering is popular for its simplicity and multiple implementations have been proposed~\cite{khalil2019tex,veedo,bentov2019tesseract,zhang2022flash,sikka,osmosis,shutternetwork,malkhi2022maximal}.

\myparagraphit{IV: MEV-aware application design.} So far the three classes of solutions are generic, but effective mitigation is possible for specific applications. Particularly interesting is the design of exchanges that are resistant to frontrunning and sandwich attacks by construction. For instance, one approach is batch auctions~\cite{budish2015high,mcmenamin2022fairtradex} (initially proposed as a countermeasure to high-frequency trading) where orders are executed in batches so that the ordering within a batch does not make a difference. 

In general, miners' tendency to maximize profits implies an  adversary model that is stronger than the honest/malicious model and the passively rational model \cite{he-htlc}. E.g., hashed timelock contract (HTLC) is widely used in payment channels and atomic swaps, but HTLC is only secure assuming honest miners because rational miners can be {\em bribed} to break the contract~\cite{tsabary2021mad}. Bribery is explicit MEV created by attackers to induce desired behaviors of miners. In \cite{he-htlc,cryptoeprint:2022/1063}, authors further showed MEV-extracting miners themselves can mount bribery attacks or collude with other participants to break the countermeasure in~\cite{tsabary2021mad}, and proposed countermeasures.

In~\cref{sec:defense}, we deep dive into the technical details of each proposed scheme and compare their goals, technical solutions, and trust assumptions. Then, we discuss how each category of ideas approaches the MEV problem and note that not all of them addressed all aspects of it. 

\parhead{Empirical study on \MEVAPs.}
Among the four categories above, \MEVAPs are the most popular MEV countermeasure in practice.
At the time of writing, more than 90\% of Ethereum
blocks are produced by \MEVAPs. 
While the working of \MEVAPs is not complex, their real-world operation is not well understood. 
In~\cref{sec:fass}, we use rich blockchain and mempool data to empirically understand the ecosystem of \MEVAPs. 

Our empirical study aims to first understand the basic structure of the \MEVAP ecosystem (e.g., what are the market shares of various \MEVAPs in terms of searcher usage and miner participation?). Further, we ask if \MEVAPs always uphold the privacy and atomicity promises. Finally, we want to understand a unique aspect of current \MEVAPs, namely their roles in  enforcing regulations. 
Recently, the Office of Foreign Assets Control (OFAC) of the US Treasury Department placed a sanction against Torando Cash~\cite{ustreasurytornado}. To be OFAC compliant, several \MEVAPs refuse to process blocks containing transactions interacting with sanctioned addresses, effectively censoring them.
\MEVAPs' involvement in enforcing US government regulation provoked heated discussion.
While the legal discussion is out of the scope of this paper, we want to understand, from a technological point of view, to what extent are \MEVAPs enforcing regulations, and the measurable implications of implementing a sanction on a blockchain. 

Our study reveals interesting findings. For example, we find that \MEVAPs do not always uphold privacy guarantees, and even supposedly private transactions fall victim to MEV attacks. This may be due to uncle blocks~\cite{unclebandit}, which highlights a tension between full privacy and short confirmation time (which is also relevant in content-agnostic ordering protocols). Moreover, \MEVAPs that claim to be compliant with OFAC regulation in fact are not strictly compliant, the evidence of which is publicly available on-chain (\cref{sec:censorship}).
We also quantify the sanction's implication on user transactions.
Since not all \MEVAPs are enforcing the sanction, it is not likely to cause transactions to be excluded. Rather, sanctioned transactions will incur a longer confirmation latency (or {\em waiting time}~\cite{eip1559}).
Using the mempool data we collected with a distributed system, we measure that sanctioned transactions on average wait for about 68\% longer than regular transactions before they can be included in a block.

\parhead{Roadmap.}
In~\cref{sec:problems}, we review the common types of MEV and the security implications. In~\cref{sec:defense}, we present the taxonomy of \numberofprojects proposed MEV countermeasures and the comparison of them.
In~\cref{sec:fass}, we report on an empirical study of the \MEVAP ecosystem. 
In~\cref{sec:related}, we discuss related works.
Finally, we end the paper with a discussion on the legal aspects of MEV and future research directions.
\section{Background: MEV and its security implications}
\label{sec:problems}

\subsection{MEV}

The term Miner/Maximal Extractable Value was first introduced by Daian et al.~\cite{daian2020flash} to refer to the value that can be extracted by a miner from manipulating the order of transactions, as an upper bound on the extractable value.
In practice, MEV extracting is a growing and lucrative industry. 
Purpose-built {\em searchers} monitor pending transactions for {\em victims} and craft MEV extraction transactions.
Crucial to the success of MEV extraction is the searcher's ability to ensure proper ordering of her transactions relative to the victim.
This can be done by setting appropriate fees or through one of the \MEVAPs (which we discuss in depth in~\cref{sec:defense} and~\cref{sec:fass}).

While being a relatively new topic, there is already substantial literature on understanding, quantifying, and mitigating MEVs~\cite{zhou2021high,eskandari2019sok,baum2021sok,wang2022impact,kulkarni2022towards,heimbach2022sok,felez2022insider,chung2022ponyta,he-htlc,qin2022quantifying,flashbotsdashboard}.
We will defer a systematic review of MEV countermeasures to~\cref{sec:defense}. Below we briefly review common MEV sources and their security implications.

\subsubsection{Common types of MEV in finance applications}

MEV may arise in various finance applications, but essentially, extracting MEV involves precisely placing  MEV-extracting transactions before, after, or around the victim.

\parhead{Frontrunning.}
Two common forms of frontrunning attacks are observed in practice.
The first form involves paying high transaction fees so that the attacker's transaction is executed before anyone else, to, e.g., take a rare market opportunity.
For example, an NFT named CryptoPunk 3860 was mistakenly listed for sale at an unusually low price. The frontrunner snatched up this valuable NFT  \footnote{\scriptsize 0xb40fd0c9a2ba2d1d5e7ee5e322f9afc5e2ec1b7e2d520b638ea83dcc9c850d02} by paying 22 ETH to the miner\footnote{\scriptsize 0xbc2cb18d0e58418d8d9c948cab319460bd409d7bd5f2978f3e52e445b351c522}.

The other form of frontrunning attack involves placing the attacker's transactions right before the victim, usually in conjunction with a subsequent backrunning to form a sandwich attack as we will discuss shortly.

\parhead{Backrunning.}
Backrunning involves placing the attacker's transaction immediately after the victim, to profit from the market dynamic created by the victim before others.

For example, when a transaction significantly (say) increases the price of a given asset in some exchange $X$, it creates an arbitrage opportunity. A backrunner can buy from another exchange $X'$ at a lower price and sell back to $X$, pocketing the difference. Note that in this case the backrunning transaction does not inflict any loss on the user, and it helps synchronize the prices between $X$ and $X'$. 

In the same vein, another example is to backrun oracle updates to take liquidation opportunities. We refer readers to~\cite{qin2021empirical} for an empirical study on liquidation.

\parhead{Sandwich attacks.} In a sandwich attack, an MEV searcher places a pair of transactions right before and after the victim's regular trade. The purpose of forming a sandwich is to manipulate asset prices so that the attacker benefits from the victim's loss~\cite{zhou2021high}.

\ignore{
Sandwich attacks are rampant on automated market makers (AMMs) such as Uniswap. For example, the victim sent a transaction \footnote{0xa2e52625b1069e03c4c5b4df00ce2ab5cd1e0430aa309a42fb0e0bf5e891d48d} to swap Ether for USDC then USDC for LEVINU on Uniswap V2. To profit from this trade, the attacker first placed a frontrunning transaction\footnote{0xfb1401f2a2ad9d1c6beac291e09145237e0919df8ccd47f1445d33c114dc7820} to swap USDC for LEVINU and then a backrunning transaction\footnote{0x5a7b8efd3e7a3d22bf6332dbf13f2fd4aba6111112b331a134054c05839658d4} to swap LEVINU for USDC, ending up with 207.64 USDC and 3,396,495.45 LEVINU more than he started out with, for \$46,822 profit after tips.
}

From the attacker's point of view, mounting sandwich attacks can be risky because if the order of the three transactions is not exactly as desired, the attacker may lose money. In practice, most sandwich attacks happen through \MEVAPs (jumping ahead, see~\cref{tab:private-mev-transactions}).

\subsubsection{Bribery attacks}
Attackers can create MEV explicitly to incentivize miners to take action in the interest of the attackers, in so-called {\em bribery} attacks. For example, a miner can be bribed to temporarily censor a transaction if the attacker sends a conflicting transaction with a higher fee~\cite{winzer2019temporary,tsabary2021mad}. 
More sophisticated bribery attacks can be facilitated by smart contracts.
The implications of bribery attacks are application-specific. In the context of payment channels and atomic swaps~\cite{he-htlc}, bribery attacks are detrimental.

\subsection{Security implications}
\label{sec:security implications}

\subsubsection{User loss}
Some MEV extraction directly causes users to lose money. 
For example, predatory sandwich attackers made a profit of more than \$3 million in November 2022 alone~\cite{sandwichprofit}, at the expense of victims.

\subsubsection{Inefficiency due to the lack of coordination}

Originally observed in~\cite{daian2020flash}, bots competing for MEV engage in on-chain bidding wars which can cause network congestion and increase transaction fees. Some MEV countermeasures can cause different forms of inefficiency. E.g., with first-come-first-served ordering, competition among MEV searchers becomes off-chain latency wars.

\subsubsection{Destabilizing consensus}
Carlsten et al.~\cite{carlsten2016instability} first showed that when transaction fees dominate block rewards, miners may deviate from honest mining and fork out high-fee blocks to attract other miners to build on the fork. MEV can be viewed as a generalized form of transaction fees paid to the miner. Having significant MEV thus exacerbates the issue. In fact, lucrative MEV extraction already dominates block rewards today~\cite{flashbotsdashboard}.
Daian et al.~\cite{daian2020flash} also described another attack vector exploiting MEV called Time-bandit attacks, which essentially augments reorg/51\% attacks with subsidy from MEV.

\subsubsection{A centralizing force}

Among many, Vitalik argued that MEV can cause centralization because there is a significant economy of scales in finding sophisticated MEV extraction opportunities~\cite{pbs-vitalik-post}.
We want to avoid a centralized and monopolized future because it harms transparency and decentralization.
Another worry is that MEV can encourage ``vertical integration''~\cite{WhyrunmevboostFlashbots} of miners and traders to form closed-door systems that harm the transparency and permissionlessness of the blockchain.
\section{MEV Countermeasures}
\label{sec:defense}

Recall that we defined four classes of MEV countermeasures in~\cref{sec:intro}. In this section, we first present a taxonomy of \numberofprojects proposed schemes from the four categories in~\cref{tab:mev-defense-summary}. Then, we compare how different classes of solutions address each aspect of the MEV problem as defined in~\cref{sec:security implications}.

\begin{table*}
    \caption{Comparison of specific systems/schemes to solve the problems caused by MEV, in four categories: {\bf I: \MEVAPs, II: Time-based ordering properties, III: Content-agnostic ordering, and IV: MEV-aware application design}.}
    \centering
        \begin{tabular}{p{3cm}|p{9cm}|p{4cm}H}
        \toprule
        \textbf{Projects/Papers} & \textbf{Goal and summary of the solution} & \textbf{Trusted parties and assumptions} & \textbf{Deployment Options} \\
        \hline
        \multicolumn{4}{c}{I: \MEVAPs} \\
        \hline
        Flashbots \cite{flashbotsauction}, Eden Network \cite{edennetworkwhitepaper}, Ethermine \cite{etherminerelay} (prior MEV-Boost) & 
        Flashbots aims to make MEV extraction easy and efficient by: 1) allowing users to specify preferred ordering and only pay if the specified ordering is satisfied, 2) Protecting the privacy of user transactions (from anyone but the trusted relay) until included on-chain, and 3) Running block space auction off-chain.
        & A single relay is trusted for privacy and respecting user-specified ordering.
        & Layer 2\\
        \hline
        MEV-Boost \cite{mevboost}
        & The goal is the same as above. The solution is similar but the centralized relay is replaced by multiple relays and builders.
        & Users choose a builder and fully trust it (including the relay it uses).
        & Currently L2 and eventually L1 (in-protocol PBS)  \\
        \hline
        Flashbots Protect~\cite{flashbotsprotect}, Ethermine RPC~\cite{etherminerelay}, Eden Network PRC~\cite{edenrpc}, bloXroute Fast Protect~\cite{bloxroutefastprotect}
        & To protect the privacy of user transactions (from anyone but the service itself) until included on-chain.
        & The service is fully trusted for privacy.
        & Layer 2\\
        \hline
        MEV Share~\cite{mevshare}, BackRunMe~\cite{backrunme}
        & Empower users to capture MEV from their transactions via a service that matches user transactions and searcher bundles using selectively disclosed information from users.
        & The service is fully trusted for privacy and transaction matching.
        & Layer 2\\
        \hline
        \multicolumn{4}{c}{II: Enforcing ordering properties} \\
        \multicolumn{4}{c}{(Notation: suppose the committee has $n$ nodes and up to $f$ are malicious)} \\
        \hline
        Aequitas \cite{kelkar2020order} & 
        block-receive-order fairness: if at least $n\gamma$ nodes receive $T$ before $T'$ for some $\frac{1}{2} < \gamma$, then $T$ should be ordered no later than $T'$.
        & $n \geq \frac{2f + 1}{2\gamma-1}$ (sync) or $n \geq \frac{4f + 1}{2\gamma-1}$ (async) 
        &  L1 (by modifying the consensus layer) or L2 (as a standalone service)\\
        \hline
        Themis \cite{kelkar2021themis} 
        & Same as above 
        & $n \geq \frac{4f + 1}{2\gamma-1}$
        & Same as above \\
        \hline 
        Pompē \cite{zhang2020byzantine} 
        & Ordering linearizability: if the highest timestamp of $T$ from all correct nodes is lower than the lowest timestamp of $T^{'}$ from all correct nodes, then $T$ is ordered before $T^{'}$.
        & $n \geq 3f + 1$ 
        & Same as above \\
        \hline 
        Quick-Fairness \cite{cachin2022quick} &
        $\kappa$-differential order-fairness:
        if the number of correct nodes who broadcast $T$ before $T^{'}$ exceeds the number of nodes who broadcast $T^{'}$ before $T$ by more than $2f + \kappa$, then  $T^{'}$ cannot be delivered before $T$ for some $\kappa \geq 0$.
        & $n \geq 3f + \kappa + 1$ 
        & Same as above \\
        \hline
        Hashgraph \cite{hashgraph}
        & Fair transaction order based on the timestamps: the fair timestamp of $T$ is the median of the times that each node claims it first received it.
        & $n \geq 3f + 1$ 
        & Same as above \\
        \hline
         Wendy \cite{kursawe2020wendy} 
        & Timed-relative fairness: if there is a time $t$ such that all honest nodes saw (according to their local clock) $T$ before time $t$ and $T^{'}$ after time $t$, then $T$ is scheduled before $T^{'}$.
        & $n \geq 3f + 1$ 
        & Layer 2  \\
        \hline 

        \multicolumn{4}{c}{III: Content-agnostic ordering} \\
        \hline
        TEX \cite{khalil2019tex}
        & Users encrypt transactions using timelock puzzles. Timelock puzzles ensure that all transactions are revealed. A similar idea is mentioned in Veedo~\cite{veedo}. 
        & The attacker cannot solve timelock puzzles much faster than honest users.
        & L1 and L2 \\
        \hline
        Tesseract \cite{bentov2019tesseract} 
        & Users encrypt transactions using keys generated in TEEs.
        & Integrity and confidentiality of TEEs
        & L1 and L2  \\
        \hline
        F3B \cite{zhang2022flash}
        & Users encrypt their transactions and store the associated secret key with the secret-management committee of $n$ trustees.
        & $n \geq 2f + 1$ for the secret-management committee
        & Layer 1 \\
        \hline
        Sikka \cite{sikka}, Osmosis\cite{osmosis}, Shutter Network \cite{shutternetwork}
        & Users threshold-encrypted transactions under a key generated by a committee of $n$ nodes. Ciphertexts are ordered using a certain policy, after which the committee threshold-decrypt and executes the transactions. 
        & Typically  $n \geq 3f + 1$
        & Layer 1 \\
        \hline
        Fino \cite{malkhi2022maximal} 
        & Fino efficiently integrates threshold encryption and secret sharing to DAG-based BFT protocol.
        & Less than $f$ malicious nodes where $n \geq 3f + 1$ (n is the number of all nodes.)
        & Layer 1 \\
        \hline
        \multicolumn{4}{c}{IV: MEV-aware application design} \\
        \hline
        CoWSwap \cite{cowswap}
        & Execute transactions in frequent batch auctions. Settlement is outsourced to solvers who compete to provide the best settlement surplus. 
        & We omit application-specific trust assumptions unless they are unique to MEV protection.
        & L1 and L2 \\
        \hline
        FairTraDEX \cite{mcmenamin2022fairtradex}
        & Frequent batch auctions realized with zero-knowledge proofs and value commitment.
        & -
        & L1 and L2\\
        \hline
        A$\rm ^2$MM \cite{zhou2021a2mm}
        & Atomically route user trades across AMMs to avoid sandwich and arbitrage opportunities.
        & -
        & L1 and L2 \\
        \hline
        P2DEX \cite{baum2021p2dex}
        & Order matching using secure multiparty computation (MPC).
        & -
        & L1 and L2 \\
        \hline
        Optimal slippage for eliminating sandwich\cite{heimbach2022eliminating}
        & Algorithmically set the slippage to balance the cost of transaction failure and that of MEV attacks.
        & -
        & \\
        \hline
        He-HTLC~\cite{he-htlc} and Rapidash~\cite{cryptoeprint:2022/1063}
        & Hashed Time-Lock Contract (HTLCs) schemes that are secure against MEV-extracting miners. Setting incentives properly so that miners are incentivized to penalize deviating players, yet not to deviate by themselves. 
        & - \\
        \bottomrule
        \end{tabular}

    \label{tab:mev-defense-summary}
\end{table*}

\subsection{MEV Auction Platforms}

\begin{figure}[]
\includegraphics[width=0.5\textwidth]{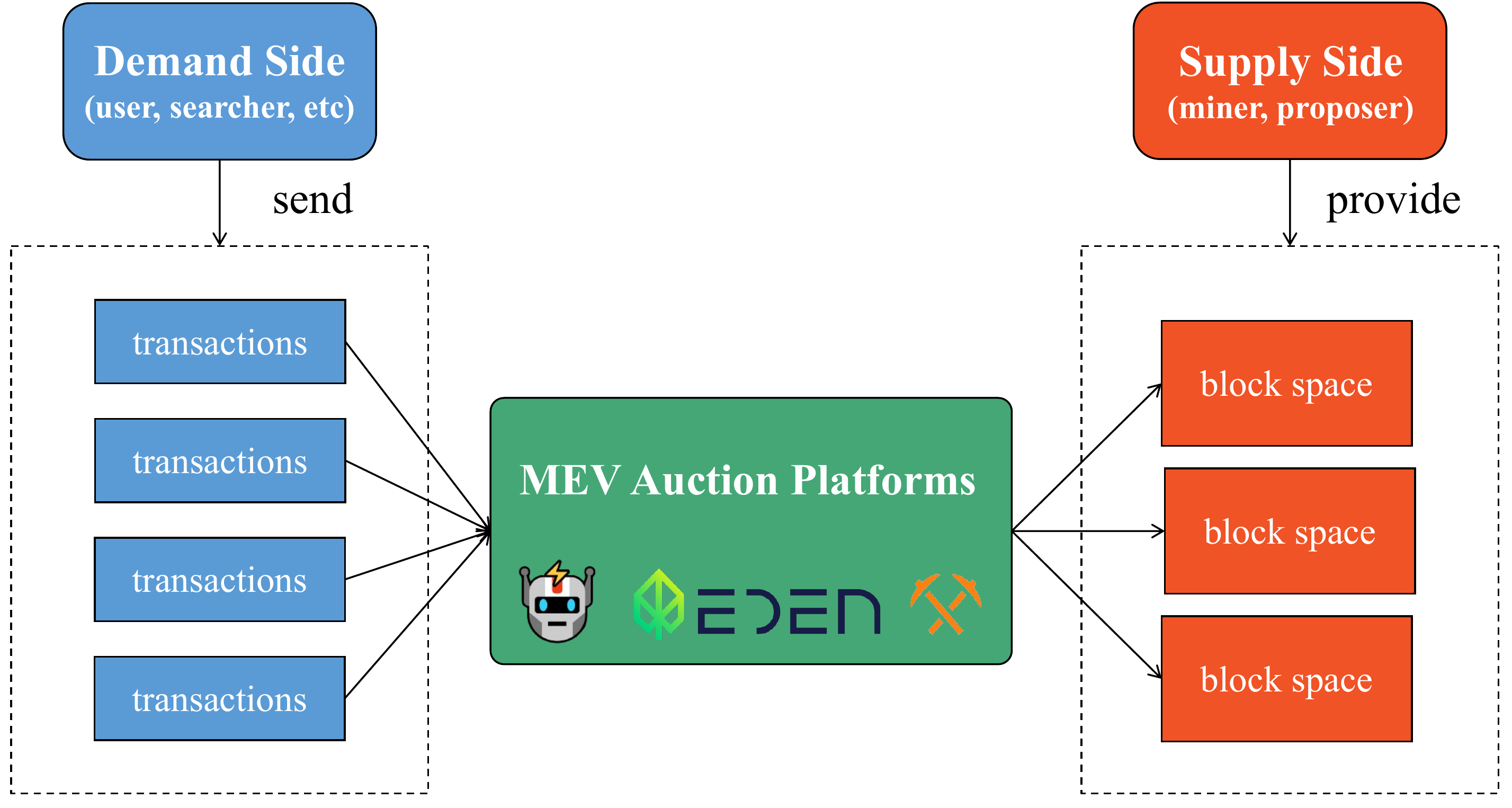}
\caption{Schematic Diagram of \MEVAPs}
\label{fig:fass-arch}
\end{figure}

The core functionality of an \MEVAP is to facilitate MEV auctions that allocate block space (sold by miners) to users (who place bids for getting their transactions included), hence the name.
\Cref{fig:fass-arch} illustrates the idea.
From a security standpoint,
\MEVAPs typically guarantee two key properties: transaction privacy (from anyone but the trusted parties) and atomicity (either the entire bundle is included in a block or none is, and the user pays only if the bundle is included).

Users of \MEVAPs can be MEV searchers and regular users. Searchers use \MEVAPs to realize their MEV-extract transactions without disclosing the transactions to other searchers and miners (otherwise they run the risk of getting frontrun by other searchers or miners).
Regular users may use \MEVAPs to hide their valuable transactions from searchers.

The recent Ethereum merge changed how \MEVAPs are implemented, so we discuss their designs separately.

\subsubsection{Pre-Merge \MEVAPs}

As the first \MEVAP, Flashbots~\cite{flashbotsauction} developed a first-price sealed-bid auction mechanism between users and miners, with the Flashbots relay as the trusted auctioneer.

A typical workflow is for users to submit a set of pre-ordered transactions (referred to as {\em bundles}) to the Flashboys relay, specifying a promised payment. Then, the relay propagates user bundles to participating miners in direct channels. Miner picks the most profitable bundles to include in their blocks. 
The relay is trusted by both users and miners:
users trust the relay to keep their transactions private and not extract MEV from them; miners trust the relay to not steal profits.

Eden Network~\cite{edennetworkwhitepaper} is a similar \MEVAP with a few key differences.
\ignore{Eden Network divides block spaces into three parts, and each part has different rules for selling block spaces so that users can purchase block spaces that meet their needs.
The head of Eden block is three priority slots, and the users can reserve these slots via a continuous auction mechanism known as a Harberger tax \cite{Harberger1962-lr}. The next part is reserved for Flashbots bundles, and these bundles will be ordered by the same rules of Flashbots Auction. The rest of the block is for users who stake EDEN in exchange for transaction ordering priority and these transactions will be ordered first by stake and then by gas price.
Eden rewards miners with EDEN tokens at predetermined times in proportion to their contribution to all produced blocks.}
Ethermine uses the same auction architecture as Flashbots Auction and accepts bundles compatible with a version of Flashbots. Ethermine claims \cite{etherminerelay} it allows users to submit bundles faster than going through the Flashbots relay, by providing a direct channel to their mining nodes.

\subsubsection{Post-Merge \MEVAPs}
\label{sec:post-merge pos}

In future versions of Ethereum, MEV auctions will see native support in the form of in-protocol Proposer-Builder Separation (PBS)~\cite{vitalikproposal}. In-protocol PBS will change the Ethereum protocol so that block building and block proposing are done by different roles (in the proof-of-work version miners do both). The major benefit of PBS is that it opens up the competition for MEV extraction to parties other than miners. 

Before in-protocol PBS is implemented,
MEV-Boost \cite{mevboost} is an intermediate realization of PBS. 
The shortcoming is that MEV-Boost still relies on trusted relays, though there are multiple of them now and in principle, anyone can become a relay.

In MEV-Boost,
a builder assembles blocks with transactions it receives from users, the public mempool, as well as the ones it inserts to extract MEV. Assembled blocks are submitted to one or more relays, with promised payments to block proposers (previously known as miners). A relay propagates received blocks to listening proposers, who finally pick the most profitable one to propose. In this process, the relay and the proposer execute a commit-then-reveal protocol so that the proposer's decision only relies on the bids and other metadata associated with a built block. not the block content.
Users pick a builder to use (or become a builder) and fully trust it, including the relay it chooses.

Among previously mentioned \MEVAPs, Flashbots and Eden now run both builders and relays, and Ethermine exits the market. New entities, such as BloXroute, joined the ecosystem as builders and relays.

\subsubsection{Private channels}
\label{sec:private channel}
For users who only want to ensure the privacy of their transactions but not the ordering, most \MEVAPs provide a service called {\em private channels}. This service can be accessed via PRC endpoints, which enjoys ease of use because users can add RPC endpoints to their wallets using existing software. Examples include Flashbots Protect \cite{flashbotsprotect}, Ethermine RPC \cite{etherminerelay} (which stopped operation after the merge~\cite{themerge}), Eden Network PRC\cite{edenrpc}, and bloXroute Fast Protect \cite{bloxroutefastprotect}. Currently, the service is fully trusted for privacy.

\subsubsection{MEV redistribution}
\label{sec:mevredistribution}

The idea of MEV redistribution aims to foster a more equitable distribution of MEV, granting users the ability to capture the MEV created by their own transactions.
To attain kickback, users first submit transactions to an intermediate service provided by \MEVAPs and selectively disclose transaction information to searchers. Searchers then propose transactions for bundling with user transactions through the intermediate service. The service will optimize MEV by matching user transactions with searcher bundles, allowing users to receive a return of the MEV. Examples include MEV Share~\cite{mevshare} and BloXroute BackRunMe~\cite{backrunme}. Presently, the service is fully trusted for privacy and transaction matching.

\subsection{Time-based ordering properties}

Drastically different from \MEVAPs, the second category of solutions in~\cref{tab:mev-defense-summary} prevents order manipulation by clearly defining the properties that transaction ordering must satisfy. 
A number of papers explore the notion of time-based ordering properties, which we review below. In the next section, we review a weaker ordering property called content-agnostic ordering.

\subsubsection{Receive-order fairness}

Receive-order fairness is first proposed by Kelkar et al. in \cite{kelkar2020order}. Basically, receive-order fairness captures the intuition of first-come-first-served ordering: if at least $\gamma$-fraction nodes receive a transaction $T$ before another transaction $T'$, then $T$ should be ordered no later than $T'$. Themis \cite{kelkar2021themis} achieves the same fairness as Aequitas \cite{kelkar2020order}, but with stronger liveness and less communication complexity. $\kappa$-differential-order-fairness achieved by Quick-Fairness \cite{cachin2022quick} can also be treated as a reparameterization of batch-order-fairness as claimed in \cite{kelkar2021themis}.
\Cref{tab:mev-defense-summary} presents the technical definitions more precisely.

Chainlink Fair Sequencing Service (FSS) \cite{juels2020fair} plans to use a receive-order fairness protocol such as Aequitas~\cite{kelkar2020order}. 

\parhead{Trust assumptions.} Above protocols assume a committee of $n$ nodes with up to $f$ of them malicious. Aequitas requires $n \geq \frac{2f + 1}{2\gamma-1}$ in the synchronous setting and $n \geq \frac{4f + 1}{2\gamma-1}$ in the asynchronous setting. Themis requires $n \geq \frac{4f + 1}{2\gamma-1}$ and Quick-Fairness requires $n \geq 3f + \kappa + 1$.

\subsubsection{Relative fairness}

Receive-order fairness is defined with respect to the relative ordering of transactions. 
Another set of fairness definitions involves using absolute time. 

Wendy\cite{kursawe2020wendy} (also known as Vega \cite{vega}) proposes relative fairness: if there is a time $t$ such that all honest validators saw transaction $T$ before $t$ and another transaction $T'$ after $t$, then $T$ must be scheduled before $T'$.
Pompē \cite{zhang2020byzantine} proposes a similar notion called ordering-linearizability
Indeed, \cite{kelkar2021themis} show that both definitions can be consolidated into a single property called fair separability.

A key difference is that Pompē relaxes the requirement so that the definition is only required if both transactions are output. In other words, it is acceptable if $T'$ is output and $T$ is not, even if all honest parties receive $T$ before $T'$. While this relaxation achieves better liveness ($T'$ cannot be held by $T$ in case of network congestion), it also permits censorship.

Hashgraph \cite{hashgraph} assigns every transaction a fair timestamp, which is the median of the time each node claims to have first received it. However, \cite{kelkar2020order} gave an attack showing that median-time-based ordering is subjective to adversary manipulation by a single attacker.

\parhead{Trust assumptions.} Similar to receive-order fairness protocols, Wendy, Pompē, and Hashgraph all require a committee of $n \geq 3f + 1$ nodes.

\subsection{Content-agnostic ordering}

Content-agnostic ordering (also known as blind-order-fairness~\cite{heimbach2022sok} and casual ordering~\cite{reiter1994securely}) is somewhat of a catch-all term because it does not correspond to a specific way of determining the ordering, as long as it is determined independent of transaction content.
In practice, content-agnostic ordering is commonly realized with a commit-and-reveal protocol. Instead of sending transactions in plaintext, users send commitments along with some metadata (e.g., the transaction fee). The miner determines an ordering based on the commitments (by hiding, they do not leak information about the transaction content), then the protocol opens the commitment, and the transactions are executed.

The commit-and-reveal step can be instantiated with different primitives, such as threshold encryption, timelock encryption, and trusted execution environments (TEEs), etc.

\subsubsection{Threshold encryption}

The general setup is a key management committee of $n$ nodes with an honest majority (or super-majority). 
Users encrypt transactions under the public key of the committee, which determines the ordering of user transactions in a protocol-specific way. Then the committee threshold-decrypts the transactions and executes them.

Sikka\cite{sikka}, Osmosis\cite{osmosis}, and Shutter Network \cite{shutternetwork} are  systems that integrate threshold encryption to Ethereum (and potentially other blockchains). Meanwhile, Fino\cite{malkhi2022maximal} proposes a way to efficiently integrate threshold encryption and secret sharing with DAG-based BFT protocol.
F3B\cite{zhang2022flash} uses a secret-management committee to store encryption keys so that when the transaction has been committed by the underlying consensus layer, its content will be later revealed by a decentralized secret-management committee.

We use (a simplified description of) Shutter Network \cite{shutternetwork} to illustrate the end-to-end transaction flow.
In Shutter Network, a group of nodes (called keypers) infrequently executes a distributed key generation (DKG) protocol to generate the main public key with the corresponding secret key secret-shared across keypers. To send a transaction $T$, the user first obtains the main public key, picks a future epoch $e$ when the transaction will be decrypted, derives the epoch-$e$ public key $PK_e$, and encrypts $T$ under $PK_e$. The ciphertext $C$ is sent to a smart contract for ordering. When epoch $e$ arrives, keypers derive the epoch-$e$ secret key and decrypt $C$ off-chain and send the plaintext to an execution smart contract for execution.

\parhead{Trust assumptions.} Using threshold cryptography assumes that a threshold of nodes is honest.

\subsubsection{Time-lock encryption}

Another option to hide the content of transactions is using timelock encryption, which allows decrypting a message once a certain time has passed. 
TEX \cite{khalil2019tex}, a front-running resilient exchange, uses time-lock puzzles to automatically decrypt transactions in case users fail to open the commitment. A similar idea also appears in Veedo documents~\cite{veedo}.

\parhead{Trust assumptions.} In order to use time-lock in commit-and-reveal schemes, we need to assume that 1) one can set the time-lock parameters relatively accurately so that reveal happens roughly at the desired time, and that 2) the attacker cannot solve time-lock puzzles much faster than honest users.

\subsubsection{Trusted Execution Environments (TEEs)} TEEs are hardware-protected isolated execution environments. TEE protects the confidentiality and integrity of the data and program inside. The state-of-the-art implementation is Intel SGX~\cite{mckeen2013innovative}, and upcoming (and potentially better) implementations include Keystone~\cite{lee2019keystone}, and Nvidia H100 GPU~\cite{nvidiatee}. TEEs also support remote attestations so that a remote user can obtain hardware-generated proofs of the code running inside. 

At a high level, TEEs can take the role of key management committees in the above solutions, by generating a pair of keys inside a TEE and publishing the public key. 
TEE can be programmed so that it releases the decryption key for epoch $e$ only if the ordering of epoch $e$ has been committed to.
However, a caveat is that TEEs do not guarantee availability. Care must be taken to ensure the liveness of TEE-based protocols.

Tesseract \cite{bentov2019tesseract} is a real-time cryptocurrency exchange built on TEEs. Tesseract relies on TEE and TLS to form secure channels between users and the exchange, so user transactions are hidden from frontrunners. Although Tesseract is an off-chain exchange, the idea can be generalized to implement content-agnostic ordering for smart contracts.

\parhead{Trust assumptions.} TEE implementation achieves confidentiality and integrity.

\subsection{MEV-aware application design}

In this section, we review application-level mitigation. We focus on decentralized exchanges (DEX) because they are currently a significant source of MEV opportunities.%

\parhead{Batch auction} Frequent Batch Auctions (FBAs) \cite{budish2015high} was proposed as a response to high-frequency trading arm races. The idea essentially is to batch execution trades in discrete time intervals. Trades in the same batch are executed at the same price, thus eliminating the advantage of manipulating the ordering within a batch. 

Although proposed for traditional markets, FBAs have been applied to DEX as well. 
CowSwap \cite{cowswap} and FairTraDEX\cite{mcmenamin2022fairtradex} are two examples.
One idea new in CowSwap is they outsource the task of settling a batch to third-party solvers, who compete for submitting the best settlement that optimizes trade surplus, avoiding the reliance on a trusted third party required in the initial FBA mechanism. FairTraDEX uses cryptography (zero-knowledge protocols in particular) and incentives to perform the settlement.

\parhead{Publicly verifiable multi-party computation.}
P2DEX \cite{baum2021p2dex} proposes a decentralized exchange construction using publicly verifiable multi-party computation where orders are matched privately via MPC servers, and misbehavior can be identified by publicly verifiable proofs and punished.

\parhead{Atomic routing.} As previous work shows that sandwich attacks are not profitable if the victim's input amount remains below the minimum profitable victim input (MVI) \cite{zhou2021high}, by combining multiple AMMs, A$^2$MM~\cite{zhou2021a2mm} can aggregate the MVI thresholds among the underlying liquidity pools to reduce the risks of sandwich attacks. Moreover, atomic routing can reduce price disparity among AMMs (in a way, the arbitrage surplus is given back to the user) and thus the overhead caused by backrunning flooding as a result of the competition to extract arbitrage.%

\parhead{Optimal slippage setting.} To use AMMs, users set a slippage to tolerate unexpected price movements. Using a low slippage run the risk of transaction failures, but setting a high slippage attracts attackers to reap the difference between the slippage and the actual price (e.g., through sandwich attacks). \cite{heimbach2022eliminating} proposes an algorithm to calculate the optimal slippage that balances the cost of transaction failures and sandwich attacks.

\subsection{Comparison of different approaches}

\begin{table*}[]
    \centering
        \caption{Comparison of different approaches to the problems caused by MEV.}
    \label{tab:approaches vs problems}
    \begin{tabular}{p{2cm}|p{3cm}|p{4cm}|p{2cm}|p{4cm}}
    \toprule
            & \textbf{Preventing user loss} & \textbf{Reducing inefficiency due to lack of coordination} & \textbf{Reducing consensus destabilizing risk}s & \textbf{Reducing centralization} \\
    \midrule
    \MEVAPs 
    & Yes and No. Users can use \MEVAPs for self-protection, but attackers can also use \MEVAPs to attack.
    & Yes. Off-chain auctions can reduce network congestion caused by PGA. 
    & No. MEV is still present in blocks.
    & Facilitating MEV extraction so non-miners can extract MEV too\\
    \hline
    Time-based ordering properties &
    Yes. Ex-post order manipulation is prevented. &
    Mostly. It will obsolete on-chain bidding war but some inefficiency may be lost to off-chain latency war. &
    Yes. &
    It removes the ordering privilege from miners but it introduces a permissioned committee. \\
    \hline
    Content-agnostic ordering &
    Mostly, but metadata leakage blind frontrunning is possible. &
    Mostly, but it depends on the ordering mechanism. If transactions are ordered by  fees, then on-chain bidding wars are possible amongst blind front-runners. &
    Yes. &
    It removes the ordering privilege from miners, but protocol-specific trust assumptions may reduce the degree of decentralization of the blockchain
    \\
    \hline 
    MEV-aware application design &
    Yes, for specific applications. &
    Yes, since MEV is eliminated for the given application. &
    Yes. &
    Partially, as it removes the ordering privilege from miners for specific applications.\\
    \bottomrule
    \end{tabular}
\end{table*}

In~\cref{sec:problems}, we defined four problems that MEV may cause. \Cref{tab:approaches vs problems} summarizes how different approaches address each problem. In the interest of space, we will refer readers to the self-explanatory table.
\section{An Empirical Study on MEV Auction Platforms}
\label{sec:fass}

Practical implementation of \MEVAPs changed significantly with the Merge~\cite{themerge} (i.e., Ethereum's transition to the Proof-of-Work based consensus protocol), but \MEVAPs remain the de facto most popular MEV mitigation in both eras. 
Prior to Merge, the largest \MEVAP, Flashbots, is adopted by more than 99.9\% of Ethereum hashrate~\cite{aflashbotinthepan}. 
Post merge, more than 90\% of blocks are produced by MEV-Boost~\cite{mevboostorg}.

Our study covers both the pre-Merge \MEVAP ecosystem (Flashbots, Eden Network, Ethermine, etc) in~\cref{sec:fass before merge}, and the MEV-Boost ecosystem post-Merge in~\cref{sec:fass post merge}.

\subsection{\MEVAPs before Merge}
\label{sec:fass before merge}

As introduced in~\cref{sec:defense}, the basic functionality of a \MEVAP is to allow users (MEV searchers and regular users) to buy block space from miners (also known as block proposers in PoS Ethereum) in an efficient and trustworthy way. 
In other words, a \MEVAP can be viewed as a two-sided marketplace, with the miners supplying block space and the users demanding it. In this section, we are interested in understanding the dynamics of such markets, in particular, with the following research questions.

\begin{itemize}[leftmargin=*]
\item What is the market share of various \MEVAPs both in terms of usage and in terms of miner participation?
\item Who uses \MEVAPs and why? 
\item Do \MEVAPs always uphold privacy and atomicity guarantees? 
\end{itemize}

\parhead{Approach}
Our main vantage point to understanding \MEVAPs is {\em the flows of private transactions}.
We trace the path through which transactions land in the blockchain without appearing in the public mempool, to understand the interaction between users (the demand side) and the miners (the supply side), as well as the intermediate \MEVAPs.

The first task is to identify private transactions and trace their flow. Below we detail the data collection process.

\subsubsection{Data collection}

As summarized in~\cref{faas-dataset}, our dataset consists of several parts: general blockchain information, private transactions (identified by our modified Ethereum nodes), transactions released by Flashbots and Eden Network, MEV labels as well as Etherscan label.%
Our data covers the block range between 15,253,306 and 15,537,393 (roughly Aug 1st, 2022 to Sept 15th, 2022).

\begin{table}[!htbp]
\centering
\caption{\MEVAP Ecosystem Dataset}
\begin{tabular}{llc}
    \toprule
    \textbf{Data Type} & \textbf{Sub-Type} & \textbf{Count} \\
    \midrule
    \multirow{2}{*}{General Information} & Blocks & 284,088   \\
    & Transactions & 50,217,035  \\
    \midrule
     \multirow{4}{*}{User Tranactions} & Private Transactions & 1,222,057 \\
    & Flashbots Transactions & 1,185,957 \\
    & Flashbots Protect Transactions & 200,053 \\
    & Eden Network Transactions & 16,416 \\
    \midrule
    \multirow{3}{*}{MEV Activities} & Sandwich & 68,517 \\
    & Arbitrage & 215,703 \\
    & Liquidation &  1,378 \\
    \midrule
    \multirow{2}{*}{Labels} & Miner Labels & 127 \\
    & MEV Bot Labels & 279 \\
    \bottomrule
    \label{faas-dataset}
\end{tabular}
\end{table}

\parhead{General information.} We collected block and transaction information from standard Web3 APIs.

\parhead{Pending transactions.} We built the Mempool Guru (\href{https://mempool.guru/}{https://mempool.guru/}) system to collect and persist pending transactions as they enter the mempool. Mempool Guru is a modular system consisting of full Ethereum nodes across the globe that record pending transactions as they enter the mempool. Mempool Guru also polls pending transactions from managed RPC nodes from Infura and QuickNode. We identify private transactions by calculating the set difference between on-chain transactions and pending transactions observed by Mempool Guru. For periods of time in which our system is down (from block 15413796 to 15414005), we use private transaction labels from Zeromev \cite{zeromev} as a supplement.

\parhead{Flashbots transactions} User transactions submitted to Flashbots are released through public APIs~\cite{flashbotsblocksapi}, from which we obtained 1,185,957 transaction hashes included in the block range of our study. We also query the private transaction status API \cite{flashbotsprotect} to get the list of 200,053 transaction hashes submitted to Flashbots Protect RPC.

\parhead{Eden network transactions.} Eden Network does not fully release user transactions as Flashbots, but by scraping the Eden Network Explorer \cite{edennetworkexplorer}, we managed to download all transactions submitted to Eden RPC in the block range of our study.

\parhead{MEV labels.} We use data from EigenPhi \cite{eigenphi} to identify MEV activities\footnote{queried on Dec 6th, 2022}. During the block range of our study, we identify 205,551 sandwich transactions (i.e., 68,517 sandwiches), 215,703 arbitrage transactions, and 1,378 liquidation transactions.

\parhead{Etherscan labels.} To identify different entities, we obtain miner and MEV Bot addresses and their related labels from Etherscan \cite{etherscanlabel}. For entities without a label, we use their addresses as identifiers.

\subsubsection{Tracing private transaction flow}

Identifying which \MEVAPs a given private transaction was submitted to is not obvious, because \MEVAPs (other than Flashbots) do not release such data in full. With partial data that is available, we attribute private transactions to \MEVAPs as follows. 
First, since Flashbots publishes all transactions it received, we can accurately attribute a private transaction to Flashbots. Eden network publishes a subset of transactions they receive, which allows us to attribute some transactions to Eden.
For transactions we cannot attribute thus far, we try to attribute them to a \MEVAP by the miner identity.
If a given private transaction is not sent through Flashbots or Eden but is mined by Ethermine, then we mark it as ``probably Ethermine'', meaning it is highly likely that the transaction was sent to Ethermine RPC in private. Finally, for each private transaction still cannot be classified, if it is mined by a participant in the Eden Network, we mark it as ``probably Eden Network'', which means it is likely sent to the Eden \MEVAP in private, but it might also be possible that the transaction was sent to {\em miners} in private. Otherwise, we consider it belongs to an unknown \MEVAP. We exclude miner payout transactions using unknown platforms because they are most likely inserted by the miner itself.

\begin{figure*}[!htbp]
\centerline{
\includegraphics[width=0.95\paperwidth]{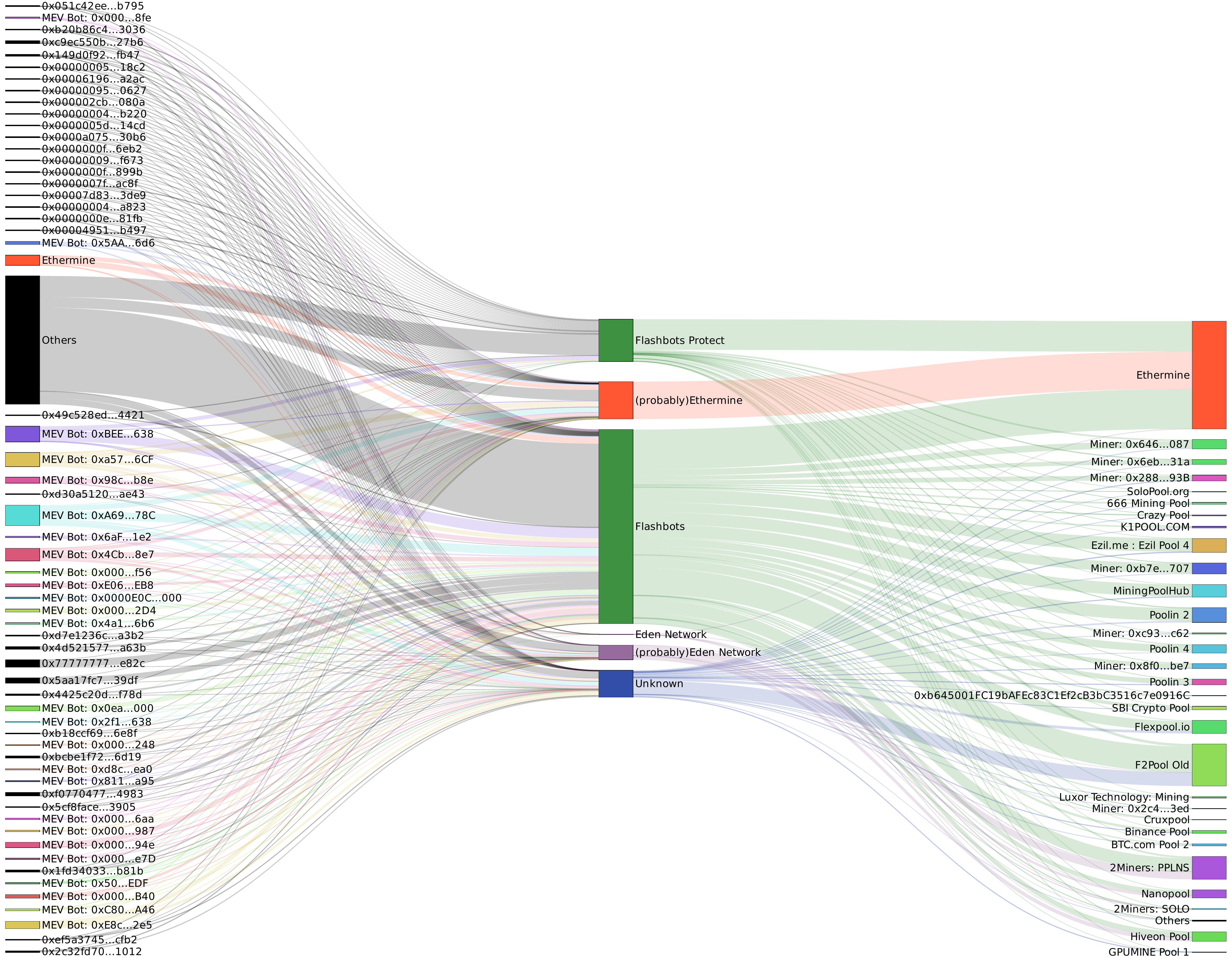}
}
\caption{Private transaction flow in the \MEVAP ecosystem (before Merge)}
\label{fig:fass-sankey}
\end{figure*}

\Cref{fig:fass-sankey} plots the private transaction flows. Each transaction starts from the sender of the private transaction from the left (identified by public keys or labels if known), goes to one of the five possible \MEVAPs (including unknown and uncertain), and finally ends at a miner on the right.
The Sankey graph provides a panoramic view of the \MEVAP ecosystem and we make several observations.

\parhead{Market shares of \MEVAPs}
As shown in \Cref{fig:fass-sankey}, Flashbots is the most popular \MEVAP in the period of our study, both in terms of usage (63.53\% private transactions are sent to Flashbots) and miner participation (56.28\% of hash power participates in Flashbots). At the same time, quite a few transactions (22.37\%) are sent through unknown \MEVAPs.
Despite the dominance of Flashbots, we note that users do leverage the diversity of \MEVAPs. More than 17.6\% of MEV auction platform users (including searchers and regular users) made use of more than two platforms.

For the market share of miners, Ethermine sold the most block spaces compared to other miners through Ethermine and Flashbots.
We note that there are many miners selling their block space through multiple \MEVAPs, such as Hiveon Pool and F2Pool Old.
{This reveals that miners are strategic and optimize for profits. Also, \MEVAPs make it easy for miners to interoperate potentially as a means to attract miner participation. For example, Eden Network allows miners to accept bundles from the Flashbots relay and include them in blocks.} We can find that some Eden miners, such as \texttt{2Miners: PPLNS} and \texttt{Cruxpool}, sell block spaces through both Eden Network and Flashbots. 
Besides, a significant part of private transactions is sent through an unknown platform, which means miners may run ``underground'' \MEVAPs. \cite{aflashbotinthepan} made a similar observation and found two miners, Flexpool and F2Pool, participate in unidentified private \MEVAPs besides Flashbots.

\parhead{Reasons to use \MEVAPs}
We can use the flow of private transactions to understand who uses \MEVAPs and why.
Looking at the left side of \cref{fig:fass-sankey}, a significant portion (45.61\%) of private transactions whose senders are miners and MEV searchers, which is in line with findings in the previous work \cite{piet2022extracting,lyu2022empirical}: In addition to protecting transaction privacy and preventing frontrunning attacks, \MEVAPs are also used for MEV extraction and redistribution of mining revenue. 

\begin{table}[]
\centering
\caption{MEV activities in private transactions versus the public mempool.}
\resizebox{\columnwidth}{!}{%
\begin{tabular}{cccc}
   \toprule
   \textbf{MEV Type} & \textbf{Private \# (\%)} & \textbf{Public \# (\%)} & \textbf{ALL \# (\%)} \\
   \midrule
    Sandwich attacks & 64,433 (94.04) & 4,084 (5.96) & 68,517 (100.00) \\
    Arbitrage & 154,758 (71.75) & 60,945 (28.25) & 215,703 (100.00) \\
    Liquidation & 1,150 (83.45) & 228 (16.55) & 1,378 (100.00) \\
    \bottomrule
\end{tabular}
}
\label{tab:private-mev-transactions}
\end{table}

To investigate how searchers use \MEVAPs for MEV activities, we calculate the intersection between the set of private transactions and the set of MEV transactions. 
\Cref{tab:private-mev-transactions} shows the number of MEV extraction activities that happened through \MEVAPs (column Private) and through public mempool (column Public). 
It is not surprising that MEV searchers prefer \MEVAPs over the public mempool, thanks to the privacy and atomicity guarantees. More than 71\% of arbitrage transactions and 83\% of liquidation transactions are made with \MEVAPs. For sandwich attacks, the percentage of using private transactions (both the frontrunning and backrunning transactions are private) is 94.04\%, which is in line with previous work \cite{aflashbotinthepan} that only 5.6\% of sandwich attacks were carried out using the public mempool.
A significantly higher percentage of sandwich attacks is conceivable: sandwich attacks require atomicity, so using a \MEVAP is particularly important.

\parhead{Self-protection.}
Regular users and developers may use a private channel service offered by \MEVAPs for self-protection against frontrunning attacks. Flashbots Protect is a private channel (defined in~\cref{sec:private channel}) for that purpose. 
We can find that 13.39\% of private transactions flow to Flashbots Protect. 
Although it is possible that some searchers may also use Flashbots Protect instead of the more powerful Flashbots Auction, it is conceivable that some of the transactions come from regular users.
This finding shows that \MEVAPs are not exclusively used by searchers to perform MEV extraction. Regular users also leverage it to protect transaction privacy.

\parhead{Do \MEVAPs uphold their promises?}
All \MEVAPs promise that user transactions sent to their services will be private from MEV bots lurking in the public mempool. However, we observe 8 sandwich attacks whose victim transaction is a private transaction, the complete list of them is in~\cref{appendix:private-txns-under-attack}.

We take one victim transaction \footnote{\scriptsize 0xdd6e4a7ab7b2c6ad5cc7aad171ee3d601a9e6cce45a355d2baa3bca65b29e156} as an example. The searcher swapped 350,000 USDC for 197.59 Ether and paid 0.2856 ETH as the tip to the miner in this transaction.
However, another searcher captured this private transaction, packed it between a frontrunning transaction\footnote{\scriptsize 0xbbac6ef9ff32b3a2bc4ac5faa4fceaf28187f60c73155c559c106784e6dc08b5} and the backrunning transaction\footnote{\scriptsize 0x5f869f0a9d92d4a8c7cd7e5d8966d3a4429a22fd6a35e200a860a2433cb0033a}, and earned 52 USDC from this sandwich attack.

One explanation for this leakage is uncle bandit attacks \cite{unclebandit} since transactions in uncle blocks (including private transactions) will be re-broadcast to the public mempool as the block is forked out.
While \MEVAPs generally warn users of the risks of uncle blocks, this finding highlights a tension between full privacy and quick confirmation.
Although \cite{lyu2022empirical} observed private leakage as well, they did not identify specific attacks. Also, since they use Etherscan ``private transaction'' labels to identify private transactions, the leakage may be due to mislabelling. See appendix for an example \cref{fig:etherscan-wrong-label} where Etherscan mislabels the public victim of sandwich attacks as private; at one point Etherscan labels all Flashbots bundles as private. The error has been later corrected.

The other possibility for private transaction leakage is simply the malice or incompetency of \MEVAPs. 
After all, there are no technical means to verify.
This is particularly conceivable for small \MEVAPs which do not have a strong reputation.
In the list, two private transactions were not sent through any public \MEVAP known to us, which suggests there exist some non-public \MEVAPs. 
\subsection{\MEVAPs post Merge}
\label{sec:fass post merge}

As discussed in~\cref{sec:post-merge pos},
MEV-Boost is an implementation of PBS by Flashbots, acting as a sidecar for the PoS node which outsources block-building to a network of builders.

At the time of writing, more than 90\% of Ethereum blocks are produced by MEV-Boost~\cite{mevboostorg}. As of November 2022, there are eight public relays in the MEV-Boost ecosystem as summarized in~\cref{tab:mevboostrelays}. Relays adhere to the MEV-Boost Relay API specification and provide the same endpoint where block builders can submit their blocks and proposers can query potential blocks.

In this section, we first provide a summary of the current MEV-Boost ecosystem, focusing on its silent feature, the diversity of relays and builders. Then, we use data to understand the relay and builder markets. Finally, we investigate the role of \MEVAPs in prevalent censorship as a result of the recent OFAC sanction against Tornado Cash~\cite{ustreasurytornado}, in particular the effectiveness and the impact of the sanction against a decentralized system.

\parhead{Data collection.}
Part of the standard relay API is for publishing various usage data, from which we can estimate the market share of builders and relays accurately.
Specifically, we query the endpoint \texttt{ProposerPayloadsDelivered} of eight public relays to collect information about blocks built through MEV-Boost (hereafter referred to as {\em MEV-Boost blocks}), including their slot numbers, builder public keys, proposer public keys, etc. The range is from block 15537394 (the Merge, Sept 15th, 2022) to 16086233 (Nov 30th, 2022).

\subsubsection{Relay differentiation}

An interesting development of the MEV Boost ecosystem is the emergence of relays with differentiating product offerings. Based on their official documents, announcements, and previous summary\cite{mevrelaylist}, we highlight the following differentiating attributes to understand the intended usage of each relay.

\begin{itemize}[leftmargin=*]
    \item \textbf{Censorship free:} Whether the relay will refuse the propagate certain transactions and bundles (e.g., those interacting with the sanctioned addresses).
    \item \textbf{Builder permissionless:} Whether any builder can submit to the relay.
    \item \textbf{All MEV Allowed}: Whether the relay will filter out certain predatory MEV transactions such as generalized frontrunning and sandwiches.
    \item \textbf{Open Source:} Whether the implementation of the relay is open source.
    \item \textbf{Profit Sharing Model}: How are the profit distribution among the relay, builders, and proposers?
\end{itemize}

As shown in ~\cref{mevboost-relay}, the Flashbots relay is permissionless and welcomes all MEV activities, but it will refuse to propagate blocks including transactions violating OFAC sanctions. bloXroute provides three versions of relays to meet the different needs for censorship and MEV activities. Builders who want to maximize their profit or ensure OFAC compliance can choose the Max profit or Regulated relay respectively. The Ethical relay and Blocknative relay will try to filter out MEV bundles which include generalized frontrunning and sandwiching. Eden relay also guarantees that their private RPC transactions will not be frontrun \cite{mevrelaylist}.

\subsubsection{Understanding the relays and builders markets}

\begin{figure}[]
\includegraphics[width=0.45\textwidth]{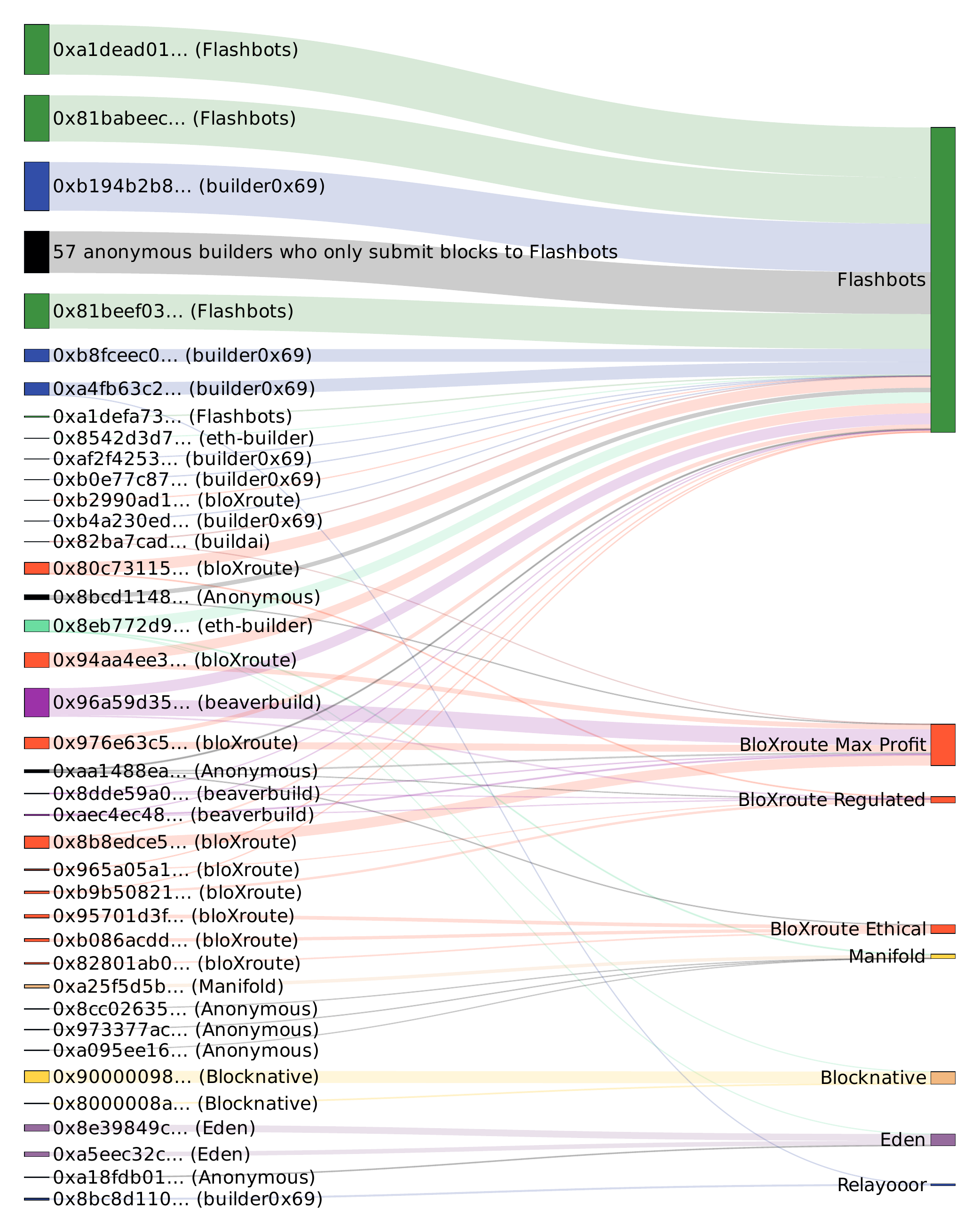}
\caption{Relationship between builders and relays (as of Nov 30th, 2022. The figure is inspired by \cite{mevboostpics} but we were able to identify 14 more builders.}
\label{fig:builder-relay-sankey}
\end{figure}

Similar to the pre-Merge analysis, we trace the flow of MEV-Boost blocks to reveal the interaction between builders and relays.
While the identity of relays is public, there is no central registrar for builders.
We need to first identify builder public keys and cluster different addresses to reduce clutter. 

\parhead{Identifying builders.}
\label{para:identifying builders}
We combine three data sources to identify builders' public keys. First, we collect the allowlist of builder public keys included in the source code of each relay. E.g., four public keys are listed in Eden's GitHub repository \cite{edenrelaygithub}.
The second source is the marks left by builders in the ``extra field'' of their blocks. For example, the mark of Flashbots builders is \texttt{``Illuminate Dmocratize Dstribute''} and the mark of bloXroute builders is  \texttt{``Powered by bloXroute''}.
Since such marks are not authenticated, we only consider a mark trustworthy if the builder keeps producing blocks with it (more than 100 blocks).
The third source is the official documents and posts published by builders and relays. E.g., we confirm that builder0x69 is related to the relay Relayooor according to their tweet\footnote{https://twitter.com/builder0x69/status/1585287608858451970}.
As shown in the table~\ref{appendix:builder_identify} in Appendix, We successfully identify 34 builders from 97 public keys.

With builder labels identified as above, we plot the block flows in~\cref{fig:builder-relay-sankey}.
The figure is inspired by \cite{mevboostpics} but we independently reproduced builder identities and we were able to identify 14 more builders (one buildai builder (0x82ba7cad), one Blocknative builder (0x8000008a), two beaverbuild builders (0x8dde59a0 and 0xaec4ec48), etc.).

 The Sankey graph allows us to understand the market shares of relays and builders and the relationship among different parties.

\parhead{Market shares of relays and builders.}
Same as before the Merge, Flashbots still currently dominate the MEV-Boost ecosystem, both in relays and builders. The Flashbots relay propagated nearly 80\% of Ethereum blocks. Again, the ecosystem evolves fast, and we refer readers to sources such as~\cite{mevboostorg} for up-to-date numbers.

Focusing on the builder side, four Flashbots builders are responsible for 33.87\% of MEV-boosted blocks as of Nov 30th, 2022. After Flashbots builders, the next biggest builder (organization) is builder0x69, which contributes 19.45\% of the MEV-Boost blocks.

\parhead{Relationship between builders and relays}
Builders are free to submit their blocks to all available relays (except that permissioned relays will refuse connections from builders not on the allowlist). We observed that builder-relay relationships could be exclusive, collaborative, or even abusive.

As shown in~\cref{fig:builder-relay-sankey}, most relays also run builders who exclusively serve the respective relays. This applies to Flashbots, Blocknative, Eden, and Manifold.  
However, we do observe cooperation among different organizations. Some BloXroute builders also use the Flashbots relay. Eth-Builder uses three relays (Eden, Blocknative, and Manifold) at the same time, and beaverbuilder uses other three relays (Flashbots, bloXroute (Max profit), and bloXroute (Regulated)).

As for builder0x69, the second largest group of builders, one of their builders only submits blocks to their own relay Relayooor, and the other builders exclusively submit to the Flashbots relay. We notice that there is a long tail of anonymous builders that submit their blocks to the Flashbots relay since it is the only permissionless relay before Relayooor and Manifold appear.

We found that malicious builders may even attack the relay for profit. Manifold, one of the permissionless relays, accepted 183 blocks from three anonymous builders and lost 7.47 ETH as a result. An official blog post~\cite{manifoldincident} explains that these anonymous builders exploited a bug to bypass Manifold's check and changed the proposed reward address. This finding shows that besides exclusive and cooperative relationships, builders may also abuse relays.

\subsubsection{Censorship in MEV-Boost ecosystem}
\label{sec:censorship}

\begin{figure}[!htbp]
	\centering
            \subfloat[Sept 15th, 2022 to Nov 7th, 2022]{
		\includegraphics[width=.95\columnwidth]{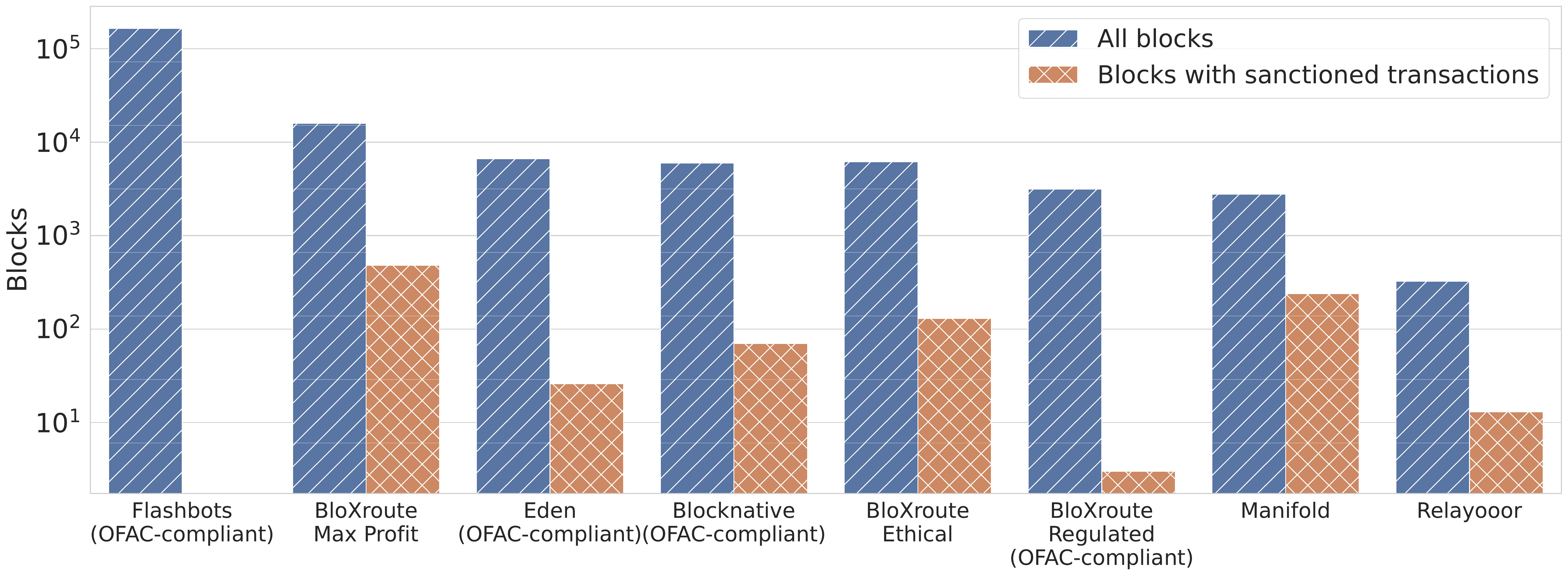}
		\label{fig:ofac-blocks-before1108}
	}
	\newline
	\subfloat[Nov 8th, 2022 to Nov 30th, 2022. The third column is the number of blocks containing transactions that interact with sanctioned addresses added on Nov 8th, 2022.]{
		\includegraphics[width=.95\columnwidth]{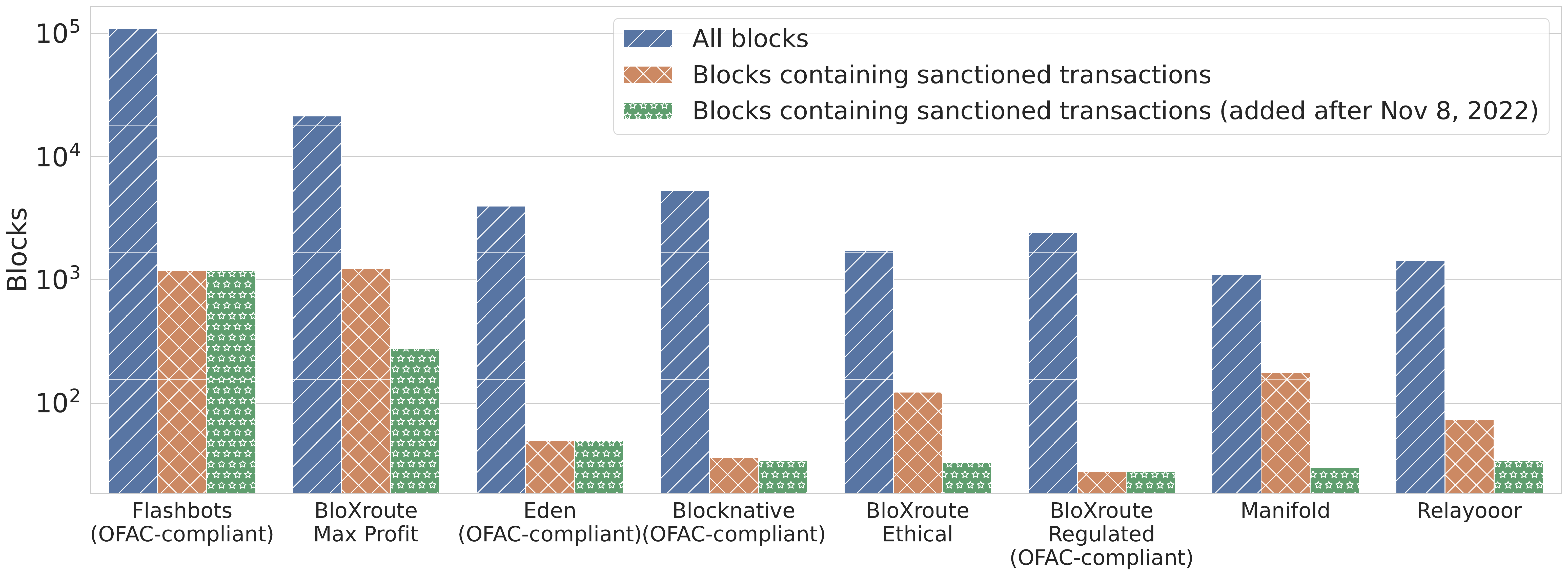}
		\label{fig:ofac-blocks-after1108}
	}
	\caption{The number of MEV-Boost blocks that include sanctioned transactions. Relays marked with OFAC-compliant are not supposed to relay blocks with sanctioned transactions.}
 \label{fig:sanction}
\end{figure}

An ongoing controversy looming over the MEV-Boost ecosystem is the ability of relays to censor transactions.
For background, the U.S. Treasury's Office of Foreign Assets Control (OFAC) recently placed a sanction on virtual currency mixer Tornado Cash \cite{ustreasurytornado}. To comply with OFAC requirements, some relays refuse to propagate transactions that interact with Tornado Cash and associated sanctioned addresses (we refer to such transactions as {\em sanctioned transactions}). 

To identify sanctioned transactions, we collect the sanctions list from \cite{ofacdownloads} and extract sanction addresses. A transaction is said to be sanctioned if 1) the sender or receiver is a sanctioned address, or 2) the execution calls a contract at a sanctioned address, or transfers ETH to a sanctioned address. We obtain transaction execution data from Etherscan~\cite{etherscanapi} to identify sanctioned transactions.

As the story is actively unfolding, we leave a comprehensive analysis for future work. 
However, we do note two interesting points. 

\parhead{Are relays actually compliant?}
As shown in~\cref{tab:mevboostrelays}, multiple relays claim to be OFAC-compliant. However, the compliance status we found does not seem to match their claims.

\Cref{fig:sanction} plots the total number of blocks relayed by each relay and the number of blocks that include sanctioned transactions, before and after November 8th, 2022. (The significance of this date will become clear momentarily.)

Interestingly, \cref{fig:sanction} shows that {\em only} Flashbots is fully OFAC compliant before Nov 8th, 2022. While the other three relays, bloXroute (Regulated), Blocknative, and Eden, who claim to be OFAC compliant, still proposed blocks containing sanctioned transactions.

On Nov 8th, 2022, OFAC updated the sanctions list, but all of the OFAC-compliant relays failed to promptly adapt to the new list. The third column in~\cref{fig:ofac-blocks-after1108} shows the number of blocks containing transactions that interact with newly sanctioned addresses. 
Notably, Flashbots has been compliant until then, but it started to accept newly sanctioned transactions after the update by OFAC.
As one example, address \texttt{0x77777FeDdddFfC19Ff86DB637967013e6C6A116C} (Tornado.Cash: TORN Token) was added to the sanctions list on Nov 8th, 2022~\cite{ofacsdnlistupdate}.
As of the timing of this paper (Nov 30th, 2022), transactions to this addressed are still accepted by relays. \texttt{0x0e399228c201352c27d7becee194a46ef2505b\\3f254470cc6afc479d467b8700} is sent and included on Nov 29th, 2022 in a block relayed by Flashbots.

It is conceivable that keeping constant track of OFAC updates is a burden for \MEVAP maintainers and human errors are possible. 
However, once any oversight occurs, it would become impossible to rectify retroactively since blockchain is immutable.
While we gave an example with Flashbots, all relays have accepted sanctioned transactions as shown in~\cref{fig:ofac-blocks-after1108}.
The full list of non-compliant transactions has been published at \url{https://tinyurl.com/ofac-non-compliant-txs}.

\parhead{What is the impact of the sanction?}
Due to the decentralized nature of blockchains, it is even unclear what a sanction is supposed to achieve on a blockchain. However, we are still able to measure what it {\em did} achieve. In short, since not all relays are censoring sanctioned transactions, the implication of the sanction is not transactions being excluded, but them being delayed. 

\begin{figure}
    \centering
    \subfloat[Cumulative Distribution]{
		\includegraphics[width=.95\columnwidth]{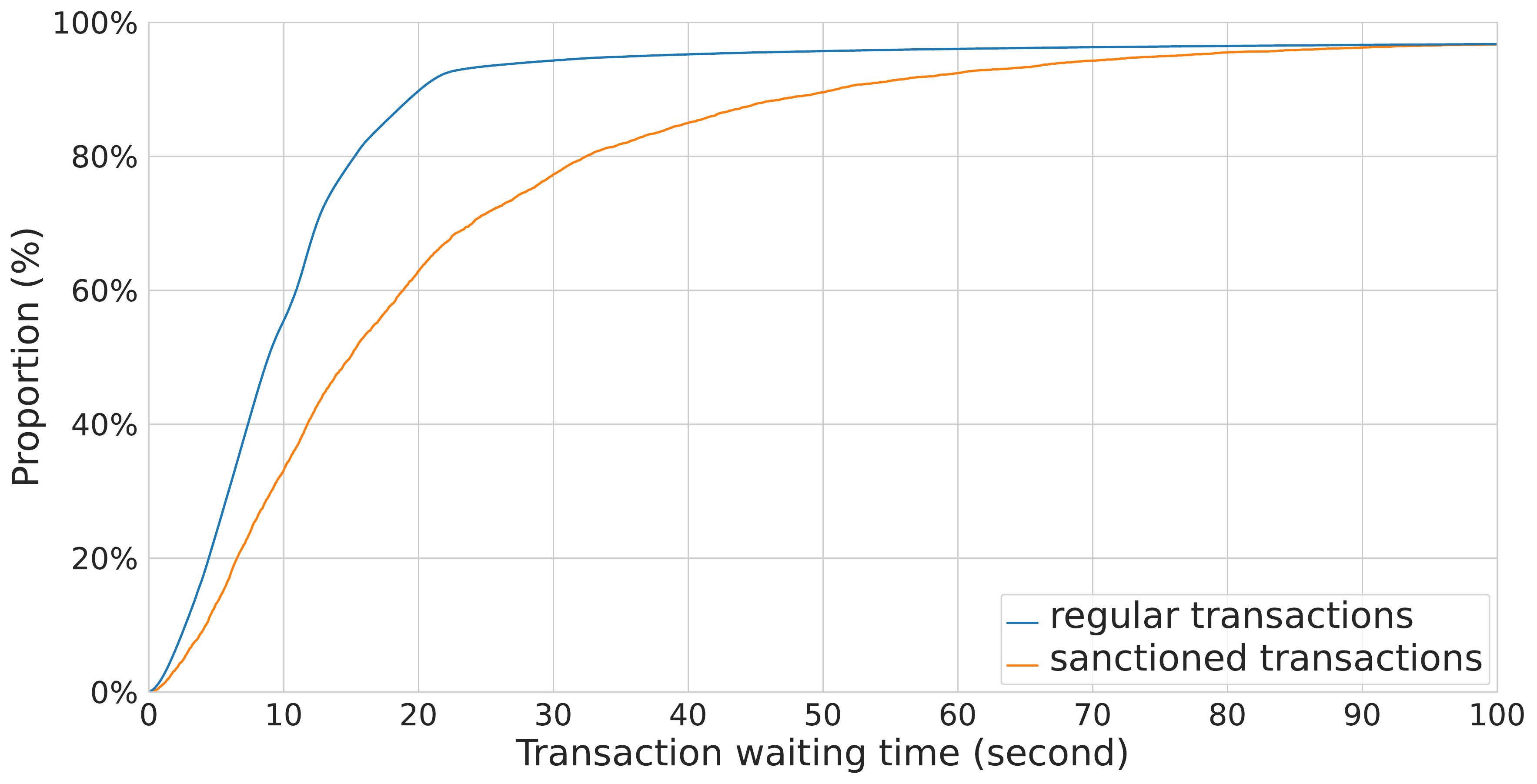}
		\label{fig:waiting-time-cdf}
	}
	\newline
	\subfloat[Probability Density]{
		\includegraphics[width=.95\columnwidth]{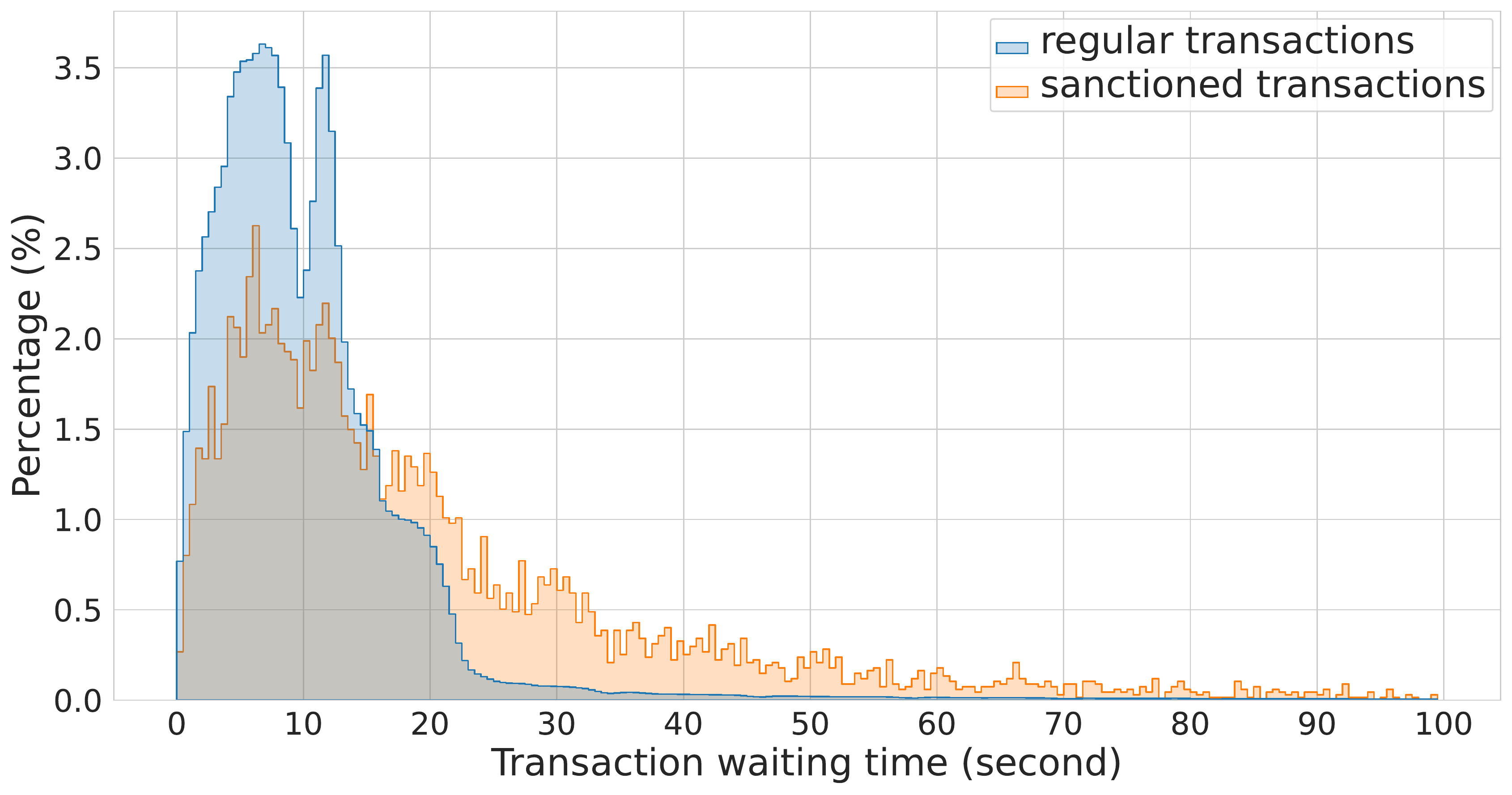}
		\label{fig:waiting-time-pdf}
	}
    \caption{Distribution of waiting time for regular transactions and sanctioned transactions. (Sept 15th, 2022 to Nov 30th, 2022)}
    \label{fig:waiting-time-distribution}
\end{figure}

To illustrate the point, we compare the waiting time~\cite{eip1559} of sanctioned transactions and regular ones. The waiting time of a given transaction is the time between when it first appears in the mempool and when it is minded in a block.  
In \cref{fig:waiting-time-distribution}, we plot the distribution of waiting time for regular transactions and sanctioned transactions. 
The median waiting times of regular transactions and sanctioned transactions are 8.87s and 14.93s respectively, thus  sanctioned transactions have to wait for about 68\% longer on average than regular transactions before they can be included in a block.

More sophisticated models can be developed to, e.g., take other factors that might affect waiting time into consideration, but the above finding shows there is a significant difference between sanctioned transactions and non-sanctioned ones, potentially caused by the sanction.
\section{Related works}
\label{sec:related}

\parhead{Study on private transactions.} 
Multiple works studied private transactions in Ethereum.
Qin et al. \cite[Appendix C]{qin2022quantifying} collected privately mined transactions prior to the emergence of \MEVAPs. Their analysis focused on mining pools' engagement in privately mining transactions.
Piet et al. \cite{piet2022extracting} studied MEV extraction activities in private transactions, and tested if Flashbots indeed mitigates negative externality. They found that 91.5\% of MEV activities are performed in private transactions (a conclusion we support) and 65.9\% of MEV profits are made by miners. Capponi et al. \cite{capponi2022evolution} drew a similar conclusion through a game theoretic analysis with empirical data support. Though it is unclear if the conclusions are robust after recent development in \MEVAPs.
Weintraub et al. \cite{aflashbotinthepan} compared MEV extraction before and after the introduction of Flashbots. They show Flashbots disproportionately benefit miners at the expense of non-miners (i.e., searchers). They also identified unknown private pools besides Flashbots (\cite[Section 6]{aflashbotinthepan}), but they did not study the \MEVAP ecosystem as a whole. They found Flexpool and F2Pool participate in other private pools besides Flashbots, consistent with our findings. Lyu et al. \cite{lyu2022empirical} analyzed private transactions on their characteristics, transaction costs, miner profits, and security impacts. \cite{lyu2022empirical} reported private transactions leakage, though they use Etherscan labels to identify private transactions, the soundness of which is unclear.

These works focus on the implications of private transactions, but we use private transactions to understand the \MEVAP ecosystem, including the market shares of various platforms, the relationship amongst different parties, and whether they uphold the security guarantees.

\parhead{SoKs on related topics.} 
Eskandari et al. \cite{eskandari2019sok} are the first to propose a taxonomy of frontrunning attacks and map the possible mitigation to front-running attacks into three categories (transaction sequencing, confidentiality, and design practices). Similarly, Baum et al. \cite{baum2021sok} assess three front-running mitigation categories (fair ordering, batching of blinded inputs, and private user balances \& secret input store) according to adversarial power of manipulating transaction order and inferring user intense. They only focus on frontrunning mitigation, which is a subset of MEV countermeasures as shown in ~\cref{tab:mev-defense-summary}.
Heimbach et al. \cite{heimbach2022sok} focus on categorizing transaction reordering manipulation mitigation schemes and analyzing the strengths and weaknesses of each mitigation scheme with a qualitative approach. While our work leverages both qualitative and quantitative approaches to understand MEV countermeasures as well as their real-world deployment.

MEV and \MEVAPs are important topics in SoK papers on DeFi too, though not as the main subject.
Werner et al. \cite{werner2021sok} discuss MEV in the context of DeFi. Zhou et al. \cite{zhou2022sok} modeled \MEVAPs as a key component in the network layer to evaluate and compare real-world DeFi attacks.
MEV is also considered as one of the security concerns to evaluate the AMM-based DEX in \cite{xu2021sok}.

\parhead{Quantifying MEV.}
Analyzing blockchain and mempool data to identify and quantify MEV starts with the original MEV paper~\cite{daian2020flash}. Subsequently, a line of works improves the techniques for MEV identification and quantification~\cite{aflashbotinthepan, capponi2022evolution, lyu2022empirical, piet2022extracting,torres2021frontrunner,wang2022cyclic,qin2022quantifying}. 
Besides, several online dashboards present real-time MEV analytic~including Flashbots Dashboard \cite{flashbotsdashboard}, EigenPhi \cite{eigenphi}, and ZeroMEV~\cite{zeromev}, etc.
Our works rely on external tools to get labels of MEV activities, but our focus is to understand different solutions and their effects on MEV rather than quantifying MEV.

\section{Discussion and future works}
\label{sec:future}

\parhead{Legal Implications of MEV.}
\label{sec:legal}
In traditional financial markets, regulated intermediaries must process trades in the best interest of the traders, in line with best execution rules. Activities such as front-running are illegal in most jurisdictions (see, e.g.,~\cite{markham1988front}).
However, there is currently no direct applicable security law governing blockchain transaction ordering in any jurisdiction. 

The recent Bank for International Settlements (BIS) research paper \cite{auer2022miners} has linked MEV to illegal market manipulation in traditional markets. The paper suggests that regulators must establish whether MEV is illegal and whether current insider trading provisions apply to MEV activities and also suggested that permissioned blockchains based on trusted intermediaries with publicly known identities may tackle MEV.
From the point of view of who engages in MEV extraction, while the decentralization nature of MEV may provide some level of regulatory protection, it has been argued that regulators may not agree with this thesis \cite{walch2018deconstructing}. However, since miners' identities are unknown, it may be difficult to enforce any regulatory measures.

\MEVAPs' engagement in enforcing OFAC's sanction against Tornado Cash poses interesting legal questions.
This paper examines the effectiveness of \MEVAPs' enforcement of OFAC sanctions, but further research should analyze the legal implication of violating OFAC regulation in a decentralized system and if \MEVAPs present a regulatory opportunity.

\parhead{Cross-domain MEV.}
MEV extraction is possible across multiple blockchains, Layer 2 systems, and exchanges (on-chain or off-chain). \cite{obadia2021unity} initiated the research on formalizing {\em cross-domain} MEV. However, understanding and mitigating cross-domain MEV largely remains an open problem.%

\parhead{MEV and Quant Trading.}
In the traditional financial market, quantitative trading algorithms can be leveraged for value extraction from trading activities. There are a few attempts to apply similar techniques for MEV purposes.
E.g., \cite{zhou2021just} uses Bellman-Ford-Moore and SMT solver to find optimal arbitrage paths. Exploring how one can leverage quantitative trading algorithms for MEV is an interesting direction.

\parhead{AI/ML and MEV}
Can AI and ML be used to analyze on-chain data to predict MEV events and build algorithms to profit from it? To the best of our knowledge, there are currently no published papers on applying AI/ML for MEV opportunities. Nevertheless, in the traditional stock market, AI/ML is used for trading. E.g., Light Gradient Boosting Machine (LightGBM) algorithm can be used in stock price prediction by constructing the minimum variance portfolio of the mean-variance model with Conditional Value at Risk (CVaR)~\cite{chen2020financial}.

\section*{Acknowledgements}
The development and operation of Mempool Guru is also supported by an Ethereum Academic Grant.

\bibliographystyle{IEEEtran}
\bibliography{IEEEabrv,ref}

\newpage
\section*{Appendix}

\subsection*{Full list of sandwich attacks with private transactions as victims}

\Cref{appendix:private-txns-under-attack} shows the details of eight sandwich attacks whose victim transaction is a private transaction, including their block number, MEV transaction types, which \MEVAPs they use, transaction hashes, along with from and to addresses.

\begin{table*}[!htbp]
    \centering
    \caption{Private transactions under sandwich attacks}
    \resizebox{\textwidth}{!}{%
    \begin{tabular}{llllll}
    \toprule
    \textbf{block number} & \textbf{MEV type} & \textbf{\MEVAP} & \textbf{transaction hash}
     & \textbf{from address} & \textbf{to address}  \\
    \midrule
15468530 & frontrun & Ethermine & 0x98ec45c7dc3808a773ab220d7aae416c967281dd7f5e0e0eaf20e2b9cadaf297 & 0x7944e84d18803f926743fa56fb7a9bb9ba5f5f24 & 0xe8c060f8052e07423f71d445277c61ac5138a2e5\\
15468530 & victim & Ethermine & 0xcb95d8a6b675dbde7237c795322bb60e4fbb32ec01dfafd983eb8f68a3952628 & 0x25173f370af28592354098a18e583f8eaa7ab264 & 0x0ef8b4525c69bfa7bdece3ab09f95bbf0944b783\\
15468530 & backrun & Ethermine & 0x75eb85a97f273b017f86215072fe5408d817d803eab63687255343f67c60bdbd & 0x7944e84d18803f926743fa56fb7a9bb9ba5f5f24 & 0xe8c060f8052e07423f71d445277c61ac5138a2e5\\
\midrule
15442739 & frontrun & unknown & 0xa4ef3607a0cc57e8275a4100147cdcf43d001e35117a148c55f19f8627cd7018 & 0xb58555fcba6479fced7de1485eb054943a09af7b & 0x00000000003b3cc22af3ae1eac0440bcee416b40\\
15442739 & victim & unknown & 0xd7555c744a92834d6210a73b4846d9f7e2f6d1de718260ce41e69d218365c821 & 0xa30622a0061513b1b6971307b66e8ba4c4f52f3c & 0x0000000000007f150bd6f54c40a34d7c3d5e9f56\\
15442739 & backrun & unknown & 0x31210110d3c21e136f7e770e2e25375747810062f624f5506718c3199ab7e093 & 0xb58555fcba6479fced7de1485eb054943a09af7b & 0x00000000003b3cc22af3ae1eac0440bcee416b40\\
\midrule
15436306 & frontrun & Flashbots & 0xd3dbbff16a1bbee3f81eb7ca595dd5c2f78ebbc97f443c99629ac6dd65fb32ec & 0xb58555fcba6479fced7de1485eb054943a09af7b & 0x00000000003b3cc22af3ae1eac0440bcee416b40\\
15436306 & victim & Flashbots & 0x7b679ef1e3d9c1ecb2930f885d039d08c9e0f8ed9b5502a3ffeb19de425a95d0 & 0xeee3109f51f0eac5212574634df62e997e550e19 & 0x0ef8b4525c69bfa7bdece3ab09f95bbf0944b783\\
15436306 & backrun & Flashbots & 0x237443460235066133938ae559614507ebf5de1ea5ec26a005276fc75fbe428b & 0xb58555fcba6479fced7de1485eb054943a09af7b & 0x00000000003b3cc22af3ae1eac0440bcee416b40\\
\midrule
15383442 & frontrun & Flashbots & 0xce2a15c9eead75e8d90d8c16a3781686f4bfdf04d15f809dc7fed19f14638434 & 0x4970197593ef5aed9d2c33409b953f5f9bb22563 & 0x00000000008c4fb1c916e0c88fd4cc402d935e7d\\
15383442 & victim & Flashbots & 0x9d287971021c5b93756e446a7962dedf5203728ca9552de704c9169e48e2cf71 & 0x16745d124412d0d3c2a83ee82ced2d93c7b4c660 & 0x2f1d79860cf6ea3f4b3b734153b52815773c0638\\
15383442 & backrun & Flashbots & 0x6b59e2b188b7392375ce6e93573486630946ca2b9fa1d6ffbd98c4cb36c54d68 & 0x4970197593ef5aed9d2c33409b953f5f9bb22563 & 0x00000000008c4fb1c916e0c88fd4cc402d935e7d\\
\midrule
15368962 & frontrun & unknown & 0xc8e441cdcec3f71a46484520eebc504c26137bede6d88313b5c89f780c0ebcf3 & 0xe2ca3167b89b8cf680d63b06e8aeefc5e4ebe907 & 0xe8c060f8052e07423f71d445277c61ac5138a2e5\\
15368962 & victim & unknown & 0x2d3358b38672b23b575c751e2f5629c703ebc06863f769c69df20dc2670b838c & 0x0cac3d1a887206e0f6169222c4504301a8b4b993 & 0xa57bd00134b2850b2a1c55860c9e9ea100fdd6cf\\
15368962 & backrun & unknown & 0x6a74371bb1fb15bb1b2ce73492cc3d83ce48cb175e3d344fb045f964f81e5d98 & 0xe2ca3167b89b8cf680d63b06e8aeefc5e4ebe907 & 0xe8c060f8052e07423f71d445277c61ac5138a2e5\\
\midrule
15350565 & frontrun & Flashbots & 0x97c0628e13773f9565102f2ff447b5d89064020bdfbd2ddfd8ab15ea5b292881 & 0x38563699560e4512c7574c8cc5cf89fd43923bca & 0x000000000035b5e5ad9019092c665357240f594e\\
15350565 & victim & Flashbots & 0x3f161a35c38ee1b5e20ef8f2fc3843e747456f3f12877a4c6b9f9319a79c374b & 0x000000cea33e55d04fb10a1af69efeb8f7e6c7f2 & 0x9cdc00c3cf228100674e4d0000e732f78d004320\\
15350565 & backrun & Flashbots & 0xb71089ecdd3276e7bf7e3b06857c4dfaf051490facee91c55d6f645a2b37305b & 0x38563699560e4512c7574c8cc5cf89fd43923bca & 0x000000000035b5e5ad9019092c665357240f594e\\
\midrule
15314350 & frontrun & Flashbots & 0xbbac6ef9ff32b3a2bc4ac5faa4fceaf28187f60c73155c559c106784e6dc08b5 & 0x36baf0d6c97efd5fd6ae995d760a84f936078759 & 0x00000000008c4fb1c916e0c88fd4cc402d935e7d\\
15314350 & victim & Flashbots & 0xdd6e4a7ab7b2c6ad5cc7aad171ee3d601a9e6cce45a355d2baa3bca65b29e156 & 0xb52a2753f420d7ad2a6588008d722b1679fad331 & 0x2f1d79860cf6ea3f4b3b734153b52815773c0638\\
15314350 & backrun & Flashbots & 0x5f869f0a9d92d4a8c7cd7e5d8966d3a4429a22fd6a35e200a860a2433cb0033a & 0x36baf0d6c97efd5fd6ae995d760a84f936078759 & 0x00000000008c4fb1c916e0c88fd4cc402d935e7d\\
\midrule
15260256 & frontrun & Flashbots & 0x5117c46039e5bbf3cecb4b941ec09415260db3bc50c7b655b4f64ca386ef21fc & 0x7944e84d18803f926743fa56fb7a9bb9ba5f5f24 & 0xe8c060f8052e07423f71d445277c61ac5138a2e5\\
15260256 & victim & Flashbots & 0xd2c01976298422bd56a3430ddbc9ebc2f9aca06a387b76af3dc75d7f0acb4493 & 0xd03154dbc4ae6beafa79f7ae6d99c12ce58f5b64 & 0x0000000000007f150bd6f54c40a34d7c3d5e9f56\\
15260256 & backrun & Flashbots & 0x3d4e04e3d385fe22668781c631871c9ad07712201eeb910148f6a5bcedd22aaa & 0x7944e84d18803f926743fa56fb7a9bb9ba5f5f24 & 0xe8c060f8052e07423f71d445277c61ac5138a2e5\\
    \bottomrule
    \end{tabular}
    }
    \label{appendix:private-txns-under-attack}
\end{table*}

\subsection*{Example of Etherscan mislabelling}

\begin{figure}[!htbp]
    \centering
            \subfloat[Screenshot from Sept 12th, 2022]{
		\includegraphics[width=\columnwidth]{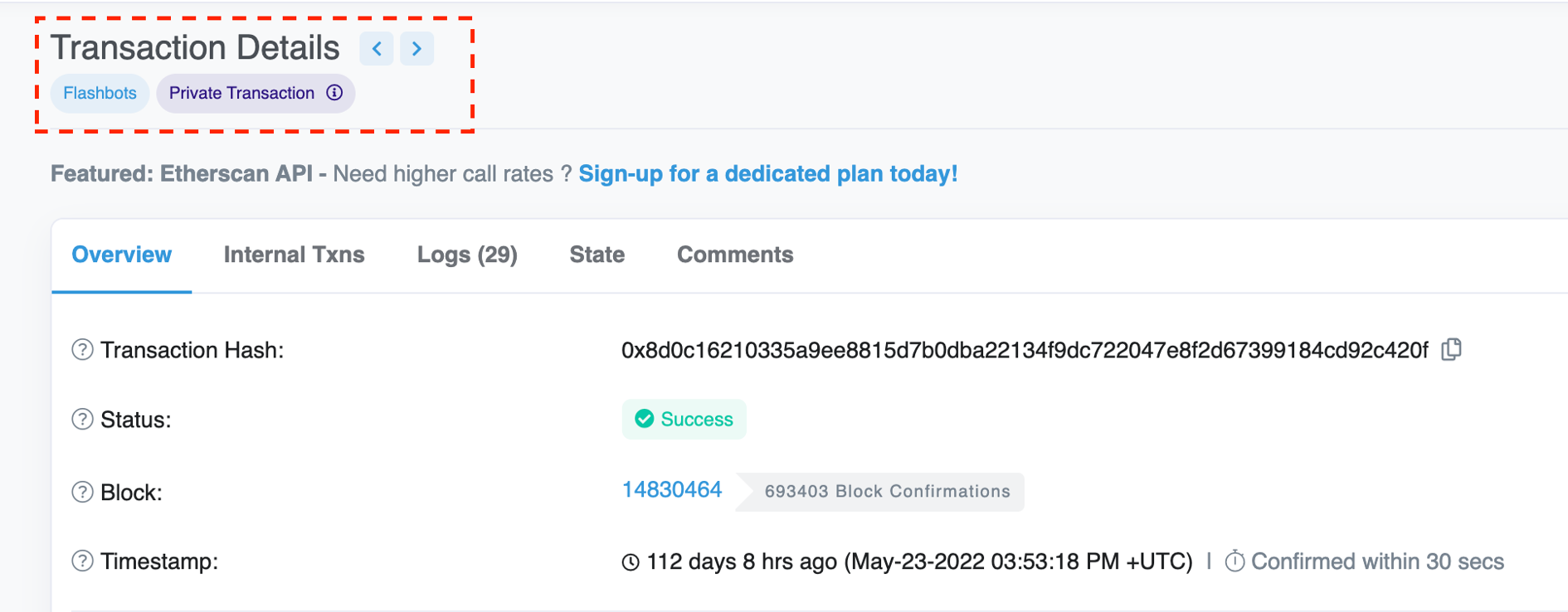}
		\label{fig:etherscan-wrong-label}
	}
	\newline
	\subfloat[Screenshot from Dec 2nd, 2022]{
		\includegraphics[width=\columnwidth]{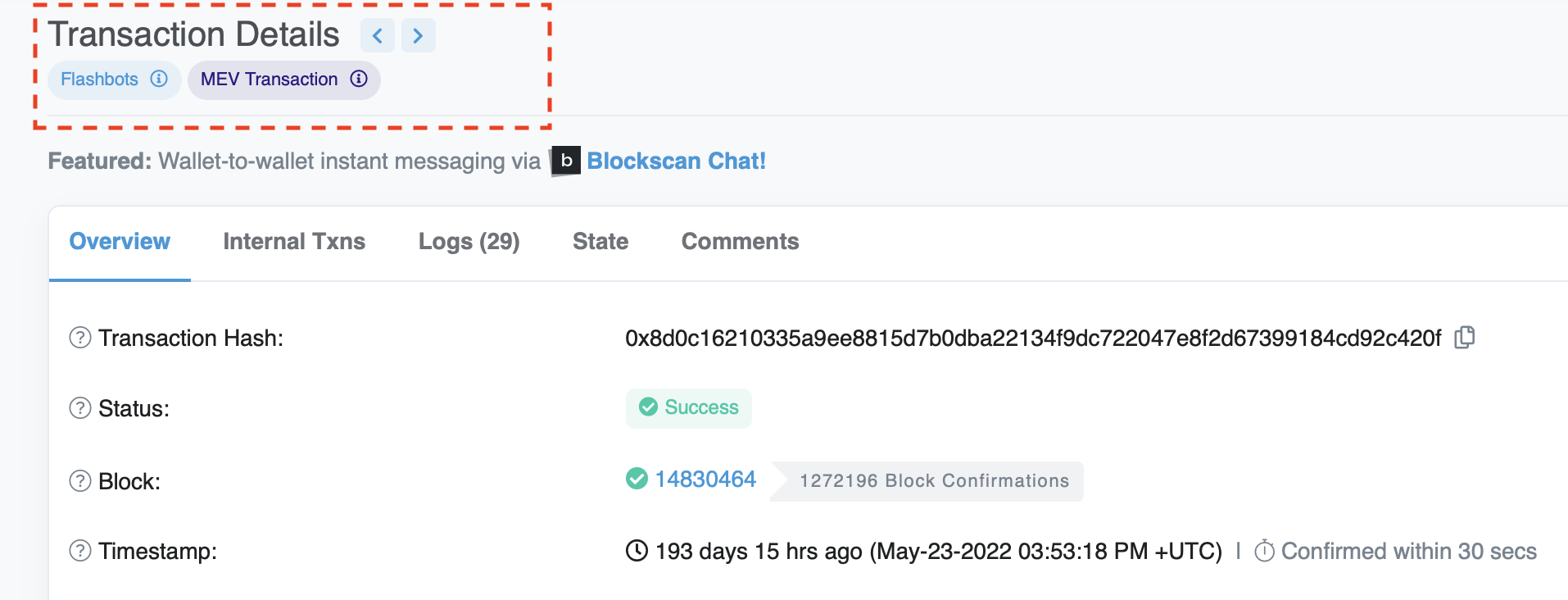}
		\label{fig:etherscan-update-label}
	}
    \caption{An example of a mislabelled private transaction in Etherscan}
    \label{fig:etherscan-label}
\end{figure}

\cref{fig:etherscan-label} gives an example that a transaction is public but mislabeled by Etherscan. From \cref{fig:etherscan-wrong-label}, we can find this transaction (whose hash is \texttt{0x8d0c16210335a9ee8815d7b0dba22134f9dc722047e8f2d67399184cd92c420f}) is labeled as a private transaction on Sept 12th, 2022. Although Etherscan had realized this problem and updated its label, as shown in \cref{fig:etherscan-update-label}.

\subsection*{Identification of MEV-Boost builders}

\cref{appendix:builder_identify} gives a complete list of the MEV-Boost builders we identify, including eight builders and 33 public keys they owned.

\begin{table}[!htbp]
    \centering
    \caption{Identified Builder}
    \resizebox{\columnwidth}{!}{%
        \begin{tabular}{lll}
        \toprule
        \textbf{Builder} & \textbf{Mark} & \textbf{Public Key (Prefix)} \\
        \midrule
        \multirow{4}{*}{Flashbots} & \multirow{4}{*}{Illuminate Dmocratize Dstribute} & 0xa1dead01... \\
        ~ & ~ & 0x81babeec... \\
        ~ & ~ & 0x81beef03... \\
        ~ & ~ & 0xa1defa73... \\
        \midrule
        \multirow{10}{*}{bloXroute} & \multirow{10}{*}{Powered by bloXroute} & 0x94aa4ee3... \\
        ~ & ~ & 0x80c73115... \\
        ~ & ~ & 0x8b8edce5... \\
        ~ & ~ & 0x95701d3f... \\
        ~ & ~ & 0xb086acdd... \\
        ~ & ~ & 0x976e63c5... \\
        ~ & ~ & 0xb9b50821... \\
        ~ & ~ & 0x82801ab0... \\
        ~ & ~ & 0xb2990ad1... \\ 
        ~ & ~ & 0x965a05a1... \\
        \midrule
        \multirow{7}{*}{builder0x69} & @builder0x69 & 0xb194b2b8... \\
        ~ & builder0x69 & 0xa4fb63c2...  \\
        ~ & by builder0x69 & 0xb8fceec0... \\
        ~ & Viva relayooor.wtf & 0x8bc8d110... \\
        ~ & by @builder0x69 & 0xb4a230ed... \\
        ~ & builder0x69 & 0xaf2f4253... \\
        ~ & by @builder0x69 & 0xb0e77c87... \\
        \midrule
        \multirow{4}{*}{Eden} & ~ & 0x8e39849c... \\
        ~ & ~ & 0xa5eec32c... \\
        ~ & ~ & 0x8931ae67... \\
        ~ & ~ & 0xb1d229d9... \\
        \midrule
        \multirow{2}{*}{Blocknative} &\multirow{2}{*}{Made on the moon by Blocknative} & 0x90000098... \\
        ~ & ~ & 0x8000008a...\\
        \midrule
        Manifold & Manifold & 0xa25f5d5b... \\
        \midrule
        \multirow{2}{*}{eth-builder} & \multirow{2}{*}{https://eth-builder.com} & 0x8eb772d9... \\
        ~ & ~ & 0x8542d3d7... \\
        \midrule
        \multirow{3}{*}{beaverbuild} & \multirow{3}{*}{beaverbuild.org} & 0x96a59d35... \\
        ~ & ~ & 0xaec4ec48... \\
        ~ & ~ & 0x8dde59a0... \\
        \midrule
        buildai & BuildAI (https://buildai.net) & 0x82ba7cad... \\
        \bottomrule
        \end{tabular}
    }
    \label{appendix:builder_identify}
\end{table}

\subsection*{Summary of MEV-Boost relay}

\cref{mevboost-relay} is a summary of eight public relays. Manifold turned to permissioned after the attack \cite{manifoldincident} and became permissionless again on Nov 21st, 2022.

\begin{table*}[!ht]
    \centering
    \caption{Public Relays}
    \label{tab:mevboostrelays}
    \begin{tabular}{cccccc}
    \toprule
        \textbf{Relay} & \textbf{Censorship Free} & \textbf{Builder Permissionless} & \textbf{All MEV Allowed} & \textbf{Open Source} & \textbf{Profit Sharing Model}  \\ 
        \midrule
        Flashbots \cite{flashbotsrelay} & \xmark & \cmark & \cmark & \cmark  & Specific to builder \\ 
        bloXroute (Max profit) \cite{bloxrouterelaymaxprofit} & \cmark & \xmark & \cmark & \xmark &  Unknown  \\ 
        bloXroute (Ethical) \cite{bloxrouterelayethical} & \cmark & \xmark & \xmark & \xmark &  Unknown \\ 
        bloXroute (Regulated) \cite{bloxrouterelayregulated} & \xmark & \xmark & \cmark & \xmark & Unknown  \\
        Blocknative \cite{blocknative} & \xmark & \xmark & \xmark & \cmark  & 100\% to validator   \\
        Eden \cite{edenrelay} & \xmark & \xmark & \xmark & \cmark  & 100\% to validator \\
        Manifold \cite{manifoldrelay} & \cmark & \cmark & \cmark & \xmark &  Specific to builder \\
        Relayooor \cite{relayooormevboostrelay} & \cmark & \cmark & \cmark & \cmark &  Specific to builder \\
        \bottomrule
    \end{tabular}
    \label{mevboost-relay}
    \end{table*}

\end{document}